\documentclass[11pt,showpacs,preprint,preprintnumbers,nofootinbib,superscriptaddress,amsmath,amssymb]{revtex4}
\usepackage[dvipdfmx]{graphicx}
\usepackage{dcolumn}
\usepackage{bm}
\usepackage{amssymb}
\usepackage{amsmath}
\usepackage{epsfig}    
\usepackage{color}
\usepackage{slashed}
\usepackage{float}
\usepackage{ulem}
\newcolumntype{I}{!{\vrule width 1.3pt}}
\begin{document} 
\title{Leptophobic $Z'$ in Models with Multiple Higgs Doublet Fields}

\author{Cheng-Wei Chiang}
\email{chengwei@ncu.edu.tw}
\affiliation{Department of Physics and Center for Mathematics and Theoretical Physics, 
National Central University, Taoyuan 32001, Taiwan}
\affiliation{Institute of Physics, Academia Sinica, Taipei 11529, Taiwan}
\affiliation{Physics Division, National Center for Theoretical Sciences, Hsinchu 30013, Taiwan}
\affiliation{Kobayashi-Maskawa Institute for the Origin of Particles and the Universe, Nagoya University, Nagoya 464-8602, Japan}
\author{Takaaki Nomura}
\email{nomura@mail.ncku.edu.tw}
\affiliation{Department of Physics, National Cheng Kung University, Tainan 70101, Taiwan}
\author{Kei Yagyu}
\email{K.Yagyu@soton.ac.uk}
\affiliation{Department of Physics and Center for Mathematics and Theoretical Physics, 
National Central University, Taoyuan 32001, Taiwan}
\affiliation{School of Physics and Astronomy, University of Southampton, Southampton, SO17 1BJ, United Kingdom}

\begin{abstract}

We study the collider phenomenology of the leptophobic $Z'$ boson from an extra $U(1)'$ gauge symmetry in models with $N$-Higgs doublet fields.  
We assume that the $Z'$ boson at tree level has (i) no $Z$-$Z'$ mixing, (ii) no interaction with the charged leptons, and (iii) no flavour-changing neutral current.  Under such a setup, it is shown that in the $N=1$ case, all the $U(1)'$ charges of left-handed quark doublets and right-handed up- and down-type quarks are required to be the same, while in the $N \ge 3$ case one can take different charges for the three types of quarks.  The $N=2$ case is not well-defined under the above three requirements. 
We study the $pp\to Z'V\to b\bar{b}V$ processes ($V=\gamma,~Z$ and $W^\pm$) with the leptonic decays of $Z$ and $W^\pm$ at the LHC.  The most promising discovery channel or the most stringent constraint on the $U(1)'$ gauge coupling constant comes from the $Z'\gamma$ process below the $t\bar{t}$ threshold and from the $t\bar{t}$ process above the threshold.  Assuming the collision energy of 8 TeV and integrated luminosity of 19.6 fb$^{-1}$, we find that the constraint from the $Z'\gamma$ search in the lower mass regime can be stronger than that from the UA2 experiment.  In the $N \ge 3$ case, we consider four benchmark points for the $Z'$ couplings with quarks.  If such a $Z'$ is discovered, a careful comparison between the $Z'\gamma$ and $Z'W$ signals is crucial to reveal the nature of $Z'$ couplings with quarks.  We also present the discovery reach of the $Z'$ boson at the 14-TeV LHC in both $N=1$ and $N\geq 3$ cases.  
\end{abstract}


\maketitle
\newpage

\section{Introduction}

After the LHC Run-I completed its operation, we have learned that there exists a scalar boson with the mass of about 125 GeV and properties consistent with the Higgs boson in the standard model (SM)~\cite{Higgs,LHC_RunI}.  This suggests that the Higgs sector consist of at least one isospin doublet scalar field.  At the same time, there has been no report about any other new particles, imposing constraints on the parameter space of new physics models, particularly their masses and couplings with the SM particles.

Among new physics models, models with an additional broken $U(1)'$ gauge symmetry provide one of the simplest framework and have been discussed based on various motivations, {\it e.g.}, grand unified theory (GUT) models~\cite{Langacker}.
This class of models features in an extra massive neutral gauge boson $Z'$ whose properties strongly depend on the nature of the $U(1)'$ symmetry.  
Therefore, detection of the $Z'$ boson and detailed measurements of its properties would be a direct probe of new physics beyond the SM. 

Searches for such $Z'$ bosons have been done in various collider experiments.  The golden search channel for models with significant $Z'$ couplings to charged leptons is the Drell-Yan (DY) process with $e^+e^- / \mu^+\mu^-$ final states.  For example, if the $Z'$ couplings to the SM fermions are exactly the same as those of the SM $Z$ boson, {\it i.e.,} the so-called sequential $Z'$ scenario, the $Z'$ mass $m_{Z'}$ is constrained to be larger than 2.86 TeV at 95\% CL using the $Z'\to e^+e^- / \mu^+\mu^-$ channels at the LHC with the collision energy of 8 TeV and integrated luminosity of 19.5 fb$^{-1}$~\cite{ATLAS_Zp_dilepton}.

However, if $Z'$ does not couple or couples very weakly with the charged leptons, the DY channel with leptonic final states is no longer useful and one has to resort to hadronic channels. 
Such a leptophobic $Z'$ boson\footnote{The collider phenomenology of a leptophobic $Z'$ boson which also couples to dark matter has been discussed 
in Ref.~\cite{Zp_DM}.   } can be realized in some GUT models, as is well-known in the $E_6$ model~\cite{lpzp_e6}. 
In this case, the most stringent experimental bound on the mass has been given by the data of $pp\to t\bar{t}$ process at the LHC when $m_{Z'}$ is above the $t\bar{t}$ threshold, with the exact value of extracted lower bound depending on the scenario.  
For example, the lower mass bound is about 1 TeV in Scenario I of the $E_6$ GUT model defined in Ref.~\cite{Chiang:2014yva}.

In our previous work~\cite{Chiang:2014yva}, we have studied the LHC phenomenology of the leptophobic $Z'$ boson inspired from the $E_6$ GUT model, concentrating on the mass regime below the $t \bar t$ threshold.  There all the $Z'$ couplings to quarks are determined uniquely according to a given embedding scheme of $E_6$~\footnote{There are six phenomenologically distinct schemes in the $E_6$ model~\cite{Rizzo,London}, all of which are examined in Ref.~\cite{Chiang:2014yva}.}.  
However, the Higgs sector, the Yukawa couplings, and their consistency with the leptophobic scenario were not discussed in that work.  
Clearly, more than one Higgs doublet field is required to construct the Yukawa Lagrangian that provides the masses of all the quarks and charged leptons if different $U(1)'$ charges are to be assigned to fermions~\cite{Ko}.

Since a SM-like Higgs boson has been discovered, we would like to examine in this work how the leptophobic condition affects the structure of the Yukawa sector and study the corresponding collider phenomenology.
More specifically, we consider models having a leptophobic $Z'$ boson associated with a $U(1)'$ gauge symmetry and $N$ Higgs doublets charged under the new symmetry.  In order to be phenomenologically viable, we require that the $Z'$ boson at tree level have~\footnote{In general, the $U(1)'$ gauge anomaly cannot be canceled within the SM particle content.  Presumably, the anomaly is canceled by introducing new particles with non-zero $U(1)'$ charges at higher mass scales, which are not discussed in this work.}: 
(i) no $Z$-$Z'$ mixing, 
(ii) no interaction with the charged leptons, and 
(iii) no flavour-changing neutral current (FCNC). 
With these conditions, we first derive consequences about the couplings of the leptophobic $Z'$ boson with quarks.  
Next, we consider constraints on the $Z'$ mass $m_{Z'}^{}$ and the $U(1)'$ charges of the quarks by using current data. 
For the mass regimes of $m_{Z'} < 2m_t$ and $m_{Z'} > 2m_t$, we take into account the constraints from $p\bar{p} \to Z' \to jj$ at the UA2 experiment~\cite{UA2} and $pp\to Z' \to t\bar{t}$ at the LHC~\cite{CMStt}, respectively.  
Finally, we perform simulations of the $pp\to Z'V\to b\bar{b}V$ ($V=\gamma,~Z$ and $W^\pm$) processes with $Z$ and $W^\pm$ decaying leptonically at the LHC for $N=1$ and $N \ge 3$ cases. 

The structure of this paper is as follows. 
In the next section, we write down the Lagrangian satisfying the above-mentioned three conditions, focusing on the Yukawa sector and deriving the $Z'$ couplings with the quarks.  The $Z' \to b\bar b$ decay branching ratio is also derived for later uses.
In Sec.~III, we discuss the constraints on the $U(1)'$ gauge coupling constant as a function of the $Z'$ mass from the UA2 experiment and the $pp\to Z' \to t\bar{t}$ process at the LHC. 
Sec.~IV shows detailed simulations of the $t$-channel $pp\to Z'V \to b\bar{b}V$ processes at the LHC.  We explicitly study the one-Higgs doublet scenario and four benchmark points for the $N \ge 3$ case.
We summarize our findings in Sec.~V.

\section{The leptophobic $Z'$ boson}

\subsection{Lagrangian}
Consider the model with a leptophobic $Z'$ boson associated with a broken $U(1)'$ gauge symmetry and $N$ Higgs doublet fields, all assumed to participate in electroweak symmetry breaking.  The most general kinetic terms, including kinetic mixing with the angle $\chi$, 
interaction terms for Higgs doublet fields $\Phi_i$ ($i=1,\dots,N$) and interaction terms for a fermion $\psi$ are given by, respectively,
\begin{align}
\mathcal{L}_{\text{kin}} &= -\frac{1}{4} W^a_{\mu \nu} W^{a \mu \nu} 
- \frac{1}{4}(\tilde{B}_{\mu\nu},\tilde{Z}_{\mu\nu}')
\left(
\begin{array}{cc}
1 & \sin\chi\\
\sin\chi & 1
\end{array}
\right)
\left(
\begin{array}{c}
\tilde{B}^{\mu\nu}\\
\tilde{Z}^{'\mu\nu}
\end{array}\right),~~  \text{with}~~X_{\mu\nu}\equiv \partial_\mu X_\nu-\partial_\nu X_\mu , \label{kin} \\
\mathcal{L}_{\text{int}}^\Phi &=\sum_{i=1}^N\left|( -ig T^a W_\mu^a -ig'Y\tilde{B}_\mu -i \tilde{g}_{Z'} \mathcal{Q}' \tilde{Z}'_\mu)\Phi_i\right|^2, \\
\mathcal{L}_{\text{int}}^\psi &= \bar{\psi} \gamma^\mu ( gT^a W^a_\mu + g' Y \tilde{B}_\mu + \tilde{g}_{Z'} \mathcal{Q}' \tilde{Z'}_\mu) \psi ~. 
\end{align}
In the above Lagrangian, 
$W_\mu^a$, $\tilde{B}_\mu$ and $\tilde{Z}'_\mu$ are the gauge fields for the $SU(2)_L$, $U(1)_Y$ and $U(1)'$ gauge groups, respectively, and the corresponding 
coupling constants (generators) are denoted by $g$ ($T^a$), $g'$ ($Y$) and $\tilde{g}_{Z'}$ (${\cal Q'}$). 
Through a non-unitary transformation, 
\begin{align}
\left(
\begin{array}{c}
\tilde{B}_\mu\\
\tilde{Z'}_\mu
\end{array}\right)=
\left(
\begin{array}{cc}
1 & -\tan\chi\\
0 & \sec\chi
\end{array}\right)
\left(
\begin{array}{c}
B_\mu\\
Z'_\mu
\end{array}\right) ~, 
\end{align}
the mixing term in ${\cal L}_{\text{kin}}$ vanishes in the basis of $Z'_\mu$ and $B_\mu$. 
However, they can still mix with each other through terms in ${\cal L}_{\text{int}}^\Phi$ after the Higgs fields develop vacuum expectation values (VEV's), as discussed below. 

In the new basis, the interaction terms are rewritten as
\begin{align}
\mathcal{L}_{\text{int}}^\Phi& = \sum_{i=1}^N \left|( -ig T^a W_\mu^a -ig'YB_\mu -ig_{Z'} \bar{\mathcal{Q}} Z'_\mu)\Phi_i\right|^2, \label{kin2} \\
\mathcal{L}_{\text{int}}^\psi& =  \bar{\psi} \gamma^\mu ( g T^a W^a_\mu + g' Y B_\mu 
+ g_{Z'} \bar{\mathcal{Q}} Z'_\mu )\psi ~, \label{int_psi}
\end{align}
where $g_{Z'}\equiv \tilde{g}_{Z'}/\cos\chi$, and 
\begin{align}
\bar{\mathcal{Q}} \equiv \mathcal{Q}'  + \delta\, Y ~,~~\text{with}~~\delta\equiv-\frac{g'}{g_{Z'}}\tan\chi ~.  
\label{qbar}
\end{align}
From Eq.~(\ref{kin2}), 
the mass terms for the neutral gauge bosons are 
\begin{align}
\mathcal{L}_{\text{mass}}
&=\frac{1}{8}(W_\mu^3,B_\mu,Z'_\mu)
\begin{pmatrix}
g^2v^2 & -gg'v^2 & -2gg_{Z'}\sum_i^N v_i^2\bar{\mathcal{Q}}(\Phi_i) \\
-gg'v^2 & g^{\prime 2}v^2 & 2g'g_{Z'}\sum_i^N v_i^2\bar{\mathcal{Q}}(\Phi_i) \\
-2gg_{Z'}\sum_i^N v_i^2\bar{\mathcal{Q}}(\Phi_i) & 2g'g_{Z'}\sum_i^N v_i^2\bar{\mathcal{Q}}(\Phi_i) 
& 4g_{Z'}^2\sum_i^N v_i^2\bar{\mathcal{Q}}^2(\Phi_i) 
\end{pmatrix}
\begin{pmatrix}
W^{3\mu} \\
B^\mu \\
Z^{\prime \mu } 
\end{pmatrix},  \notag\\
&=\frac{1}{8}(Z_\mu,A_\mu,Z'_\mu)
\begin{pmatrix}
g_Z^2v^2 & 0 & -2g_Z^{}g_{Z'}\sum_i^N v_i^2\bar{\mathcal{Q}}(\Phi_i) \\
0 &  0 & 0 \\
-2g_Z^{}g_{Z'}\sum_i^N v_i^2\bar{\mathcal{Q}}(\Phi_i) & 0 & 4g_{Z'}^2\sum_i^N v_i^2\bar{\mathcal{Q}}^2(\Phi_i) 
\end{pmatrix}
\begin{pmatrix}
Z^\mu \\
A^\mu \\
Z^{\prime \mu} 
\label{gaugemass}
\end{pmatrix}, 
\end{align}
where $v_i$ ($i=1,\dots N$) are the VEV's of $\Phi_i$ and satisfy the sum rule $\sum_{i}v_i^2 = v^2 = (\sqrt{2}G_F)^{-1}$, with $G_F$ being the Fermi decay constant and $g_Z^{}=\sqrt{g^2+g^{\prime 2}}$.  
In the last step of the above expression, the $SU(2)_L$ and $U(1)_Y$ gauge fields are rotated to the mass eigenbasis in the usual way:
\begin{align}
\begin{pmatrix}
W^3_\mu \\
B_\mu
\end{pmatrix}
=
\begin{pmatrix}
\cos\theta_W & \sin\theta_W \\
-\sin\theta_W & \cos\theta_W
\end{pmatrix}
\begin{pmatrix}
Z_\mu \\
A_\mu
\end{pmatrix},
\label{weakmixing}
\end{align}
where $\theta_W$ is the weak mixing angle. 

In this paper, we restrict our considerations to the $Z'$ boson that at tree level has
\begin{description}
\item{(I)} no mixing with the $Z$ boson; 
\item{(I\hspace{-0.5mm}I)} no interactions with left- and right-handed leptons (leptophobic condition); and 
\item{(I\hspace{-0.5mm}I\hspace{-0.5mm}I)} no FCNC via neutral scalar bosons~\footnote{Tree-level FCNC's via neutral gauge bosons are automatically forbidden by the Glashow-Iliopoulos-Maiani mechanism~\cite{GIM}. }. 
\end{description}
In Eq.~(\ref{gaugemass}), the term $\sum_{i=1}^N v_i^2 \bar{\mathcal{Q}}(\Phi_i)$ gives rise to non-zero mixing between $Z$ and $Z'$.  We therefore impose 
\begin{align}
\sum_{i=1}^N v_i^2 \bar{\mathcal{Q}}(\Phi_i) = 0 \label{nomix}
\end{align}
to satisfy (I)\footnote{This requirement can be guaranteed when the $Z'$ boson does not couple with the Higgs boson, but obtains its mass from the VEV of another scalar boson that breaks the $U(1)'$ gauge symmetry.}.
Secondly, the leptophobic condition (I\hspace{-0.5mm}I) demands
\begin{align}
 \bar{\mathcal{\mathcal{Q}}}(L_L) =  \bar{\mathcal{Q}}(e_R) = 0 ~,  \label{LPcond}
\end{align}
where $L_L$ and $e_R$ are respectively the left-handed lepton doublet field and the right-handed charged lepton field.
Finally, we consider condition (I\hspace{-0.5mm}I\hspace{-0.5mm}I). 
In general, a model with a multi-doublet structure has FCNC's at tree level because 
the fermion mass matrix may not be proportional to the corresponding Yukawa interaction matrix. 
In that case, the interaction matrix is non-diagonal in the fermion mass eigenbasis. 
To avoid such a situation, we require that each of up-type quarks, down-type quarks, and charged leptons couple to only one Higgs doublet; 
namely, the Yukawa Lagrangian in the $N$ Higgs doublet model assumes the following form: 
\begin{align}
\mathcal{L}_Y =  -Y_d \bar{Q}_L\Phi_d d_R-Y_u\bar{Q}_L\tilde{\Phi}_u u_R -  Y_e \bar{L}_L\Phi_e e_R+ \text{h.c.},\label{Yukawa_n}
\end{align}
where $\Phi_d$, $\Phi_u$ and $\Phi_e$ are same or different Higgs fields of $\Phi_i$, $Q_L$, $d_R^{}$ and $u_R^{}$ represent respectively the left-handed quark doublet field, the right-handed down-type quark field and the right-handed up-type quark field. 
For the Lagrangian in Eq.~(\ref{Yukawa_n}) to satisfy condition (I\hspace{-0.5mm}I\hspace{-0.5mm}I), 
$\bar{\cal Q}(\Phi_i)$ should be all distinct; {\it i.e.}, 
\begin{align}
\bar{\cal Q}(\Phi_i) \neq \bar{\cal Q}(\Phi_j) \qquad \mbox{for } i\neq j ~,
\label{fcnc}
\end{align} 
because both $\Phi_i$ and $\Phi_j$ could couple to the same type of fermions otherwise. 
According to Eq.~(\ref{Yukawa_n}), the $Z'$ charges of the particles have the relations:
\begin{align}
&-\mathcal{Q}'(L_L)+\mathcal{Q}'(\Phi_e)+\mathcal{Q}'(e_R)=-\bar{\mathcal{Q}}(L_L)+\bar{\mathcal{Q}}(\Phi_e)+\bar{\mathcal{Q}}(e_R)=0 ~,
\label{Ye2} \\
&-\mathcal{Q}'(Q_L)+\mathcal{Q}'(\Phi_d)+\mathcal{Q}'(d_R)=-\bar{\mathcal{Q}}(Q_L)+\bar{\mathcal{Q}}(\Phi_d)+\bar{\mathcal{Q}}(d_R)=0 ~,
\label{Yd2}\\
&-\mathcal{Q}'(Q_L)-\mathcal{Q}'(\Phi_u)+\mathcal{Q}'(u_R)=-\bar{\mathcal{Q}}(Q_L)-\bar{\mathcal{Q}}(\Phi_u)+\bar{\mathcal{Q}}(u_R)=0 ~,
\label{Yu2}
\end{align} 
where Eq.~(\ref{qbar}) has been used.  From Eqs.~(\ref{LPcond}) and (\ref{Ye2}), we have 
\begin{align}
\bar{\mathcal{Q}}(\Phi_e)=0 ~. 
\label{LPn3}
\end{align}
That is, the Higgs doublet that couples to the charged leptons cannot carry nonzero $Z'$ charge.

We now discuss consequences of Eqs.~(\ref{Yd2}) and (\ref{Yu2}) for several special cases of $N$. 
If $N=1$, corresponding to the case where $\Phi_e=\Phi_d=\Phi_u$, the $Z'$ charges of all quarks have to be the same:
\begin{align}
\bar{\mathcal{Q}}(Q_L)= \bar{\mathcal{Q}}(u_R)=\bar{\mathcal{Q}}(d_R) ~.
\label{1H_relation}
\end{align} 
In the case of $N=2$, the conditions in Eqs.~(\ref{nomix}) and (\ref{LPn3}) require 
the $Z'$ charges of the two doublets to be zero. 
However, this is in contradiction with Eq.~(\ref{fcnc}).
Therefore, the two-doublet case cannot simultaneously satisfy all the requirements (I), (I\hspace{-0.5mm}I) and (I\hspace{-0.5mm}I\hspace{-0.5mm}I). 

In the case of $N\geq 3$, 
we obtain from Eqs.~(\ref{Yd2}) and (\ref{Yu2}) that
\begin{align}
\bar{\mathcal{Q}}(\Phi_d) &= \bar{\mathcal{Q}}(Q_L)-\bar{\mathcal{Q}}(d_R) ~, 
\label{eq:chargePhi_u}
\\
\bar{\mathcal{Q}}(\Phi_u) &=-\bar{\mathcal{Q}}(Q_L)+\bar{\mathcal{Q}}(u_R) ~. 
\label{eq:chargePhi_d}
\end{align}
In addition, from Eqs.~(\ref{nomix}) and (\ref{LPn3}) we have the relationship among the Higgs VEV's and $\bar{ \cal Q}$ charges as 
\begin{align}
&\langle \Phi_{u}^0 \rangle^2 \bar{\cal Q}(\Phi_u) + \langle \Phi_{d}^0 \rangle^2 \bar{\cal Q}(\Phi_d) 
=- \sum_{i \not= u,d} \langle \Phi_i^0 \rangle^2 \bar{\cal Q}(\Phi_i) ~,
\label{vev1}
\end{align}
where the sum is over all Higgs doublets other than $\Phi_u$ and $\Phi_d$ that do not couple to fermions. 
In the case of $N=3$ in particular, Eq.~(\ref{vev1}) can be rewritten as 
\begin{align}
\frac{ \bar{\mathcal{Q}}(\Phi_d)}{\bar{\mathcal{Q}}(\Phi_u)}
= - \left(\frac{\langle \Phi_{u}^0 \rangle}{\langle \Phi_{d}^0 \rangle}\right)^2 ~. \label{chargeratio}
\end{align}
Therefore, the three-Higgs doublet case is the minimal setup that allows different $Z'$ charges for all the three types of quark fields; $i.e.,$ $\bar{{\cal Q}}(Q_L)$, $\bar{{\cal Q}}(u_R^{})$, and $\bar{{\cal Q}}(d_R^{})$ being all different.
Moreover, the two charges $\bar{\mathcal{Q}}(\Phi_d)$ and $\bar{\mathcal{Q}}(\Phi_u)$ have opposite signs according to Eq.~(\ref{chargeratio}). Also, it should be noted that if $\langle \Phi_{u}^0 \rangle$ and $\langle \Phi_{d}^0 \rangle$ are the same, $\bar{{\cal Q}}(u_R^{}) = \bar{{\cal Q}}(d_R^{})$ according to Eqs.~(\ref{eq:chargePhi_u}) and (\ref{eq:chargePhi_d}).

Even though we have explicitly imposed the condition of Eq.~(\ref{nomix}), $Z'$-$Z$ mixing~\cite{Zp_mixing,Erler} can still occur at loop levels. 
In our scenario, there are SM quark loop contributions to the two point function of $Z$-$Z'$ mixing at the one-loop level. 
Its transverse part is calculated as 
\begin{align}
\Pi_{ZZ'}(p^2) = \frac{3g_Z^{}g_{Z'}^{}}{2\pi^2}\sum_q\Bigg\{&
a_{Z,q}^{\text{SM}}a_q m_q^2\left[\int_0^1dx\ln \Delta_F - \Delta \right]\notag\\
&+p^2(v_{Z,q}^{\text{SM}}v_q + a_{Z,q}^{\text{SM}}a_q)\left[\int_0^1dx (x^2-2x+1/2)\ln \Delta_F - \frac{\Delta}{6} \right]
\Bigg\},
\end{align}
where $v_{Z,q}^{\text{SM}}$ ($a_{Z,q}^{\text{SM}}$) and  $v_q$ $(a_q)$ are respectively the vector (axial-vector) couplings of the $Zq\bar{q}$ and $Z'q\bar{q}$ vertices with corresponding expressions given in Eqs.~(\ref{eq:vqaq}) and (\ref{vsm}), $\Delta_F=-x(1-x)p^2-m_q^2$, and $\Delta$ is the divergent part of the loop integral. 
In the $\overline{\text{MS}}$ scheme, $\Delta$ is simply replaced by $\ln \mu^2$ with $\mu$ being an arbitrary mass scale. 
Such one-loop contributions give rise to nonzero off-diagonal (1,3) and (3,1) elements of the mass matrix  Eq.~(\ref{gaugemass}). 
In this case, the $Z$-$Z'$ mixing can be calculated for given values of momentum $p^2$ and scale $\mu$ as
\begin{align}
\tan2\theta_{ZZ'} \simeq \frac{2\Pi_{ZZ'}(p^2)}{m_Z^2-m_{Z'}^2} 
+ {\cal O}\left(
\frac{[\Pi_{ZZ'}(p^2)]^2}{m_Z^4},\frac{[\Pi_{ZZ'}(p^2)]^2}{m_{Z'}^4} 
\right). 
\end{align}
Taking $m_{Z'}^{}=150$ GeV, $g_{Z'}^{}=0.1$, $v_u=v_d=1$ ($v_u=v_d=0.5$) and $a_u=a_d=0$ ($a_u=a_d=0.5$), we obtain $\sin\theta_{ZZ'}=1.65~(1.04)\times 10^{-3}$ with 
$p^2=\mu^2=m_{Z'}^2$. In this calculation, we use $m_t=173.07$ GeV and $m_b=3.0$ GeV, and all the other quark masses are neglected. 
Although such a mixing effect can contribute to additional $Z'$ boson productions in collider experiments, it is negligibly small because $\theta_{ZZ'}=\mathcal{O}(10^{-3})$.

\subsection{$Z'$ couplings to quarks}

In a model with $N$ Higgs doublet fields, the $Z'$ interactions with one generation of quarks are given by Eq.~(\ref{int_psi}) as
\begin{align}
\mathcal{L}_{\text{int}}^q = g_{Z'}\Big[ \bar{\mathcal{\mathcal{Q}}}(Q_L)\bar{Q}\gamma^\mu P_L Q + 
\bar{\mathcal{\mathcal{Q}}}(u_R)\bar{u}\gamma^\mu P_R u +\bar{\mathcal{\mathcal{Q}}}(d_R)\bar{d}\gamma^\mu P_R d\Big]Z_\mu' ~, 
\label{eq:QQZpr}
\end{align}
where the projection operators $P_{L,R}=(1\mp \gamma_5)/2$. 
Alternatively, Eq.~(\ref{eq:QQZpr}) can be written in terms of the vector coupling $v_q$ and axial-vector coupling $a_q$ as
\begin{align}
\mathcal{L}_{\text{int}}^q =  g_{Z'}\bar{q}\gamma^\mu (v_q-\gamma_5 a_q)qZ_\mu' ~,
\label{eq:int2}
\end{align}
where 
\begin{align}
&v_q=\frac{1}{2}\left[\bar{\mathcal{\mathcal{Q}}}(Q_L)+\bar{\mathcal{\mathcal{Q}}}(q_R)\right],\quad
a_q=\frac{1}{2}\left[\bar{\mathcal{\mathcal{Q}}}(Q_L)-\bar{\mathcal{\mathcal{Q}}}(q_R)\right] ,\quad 
\text{for}~~q=u,\,d. \label{eq:vqaq}
\end{align}
When $Q_L$ carries a nonzero $\bar{\cal Q}$ charge, one can always normalize its value to unity by rescaling the coupling $g_{Z'}$.  Therefore, we take the value of $\bar{\mathcal{\mathcal{Q}}}(Q_L)$ to be either 1 or 0 in the following analyses. 
In this convention, $a_q$ is equal to $1-v_q$~($-v_q$) when $\bar{\mathcal{\mathcal{Q}}}(Q_L)=1~(0)$. 
In the one-Higgs doublet case, only $g_{Z'}$ is a free parameter and all the others are fixed according to Eq.~(\ref{1H_relation}) as:
\begin{align}
v_u = v_d=1,\quad a_u =a_d = 0 ~.
\end{align}
This means that the $Z'$ couplings to the quarks must be vectorial.
The interaction Lagrangian in Eq.~(\ref{eq:int2}) can also be rewritten in terms of the chiral couplings as  
\begin{align}
\mathcal{L}_{\text{int}}^q&= 
g_{Z'}\bar{q}\gamma^\mu (g_L^q\,P_L+g_R^q\,P_R)qZ_\mu' ~,
\label{eq:int3}
\end{align}
where 
\begin{align}
g_L^q=v_q+a_q=1,\quad g_R^q=v_q-a_q=2v_q-1 ~.
\label{lrcoupling}
\end{align}

\subsection{$Z'$ Decays \label{sec:ZprDecay}}

\begin{figure}[t]
\begin{center}
\includegraphics[width=80mm]{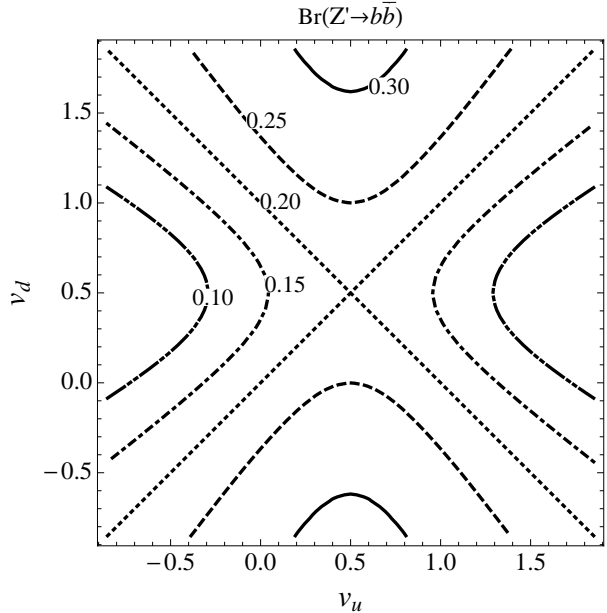}
\end{center}
\caption{Contour plot of the $Z' \to b \bar b$ branching fraction on the $v_u$-$v_d$ plane, 
assuming $m_{Z'} < 2 m_t$ and ignoring the quark masses.  
\label{Br_contour}}
\end{figure}

The partial width of the $Z'$ decaying into a pair of quarks is given by
\begin{align}
\Gamma(Z' \to q \bar q ) &= g_{Z'}^2\frac{ m_{Z'}}{4 \pi}\left[v_q^2(1+2x_q)+a_q^2(1-4x_q)\right]\sqrt{1-4x_q}~,~~
\text{with}~~x_q=\frac{m_q^2}{m_{Z'}^2} ~.
\label{eq:partialwidth} 
\end{align}
The corresponding branching fraction is 
\begin{align}
\text{Br}(Z'\to q\bar{q}) = \frac{\Gamma(Z' \to q \bar q )}{\Gamma_{Z'}} ~,
\end{align}
where the total decay width $\Gamma_{Z'} = \sum_q \Gamma(Z' \to q \bar q )$ summed over all quarks with mass less than $m_{Z'}/2$. 
When $m_{Z'} \leq 2m_t$ but much greater than $2m_b$, the branching fraction is approximately 
\begin{equation}
\text{Br}(Z' \to q \bar q) =  \frac{2(v_q-1/2)^2 + 1/2}{6(v_d-1/2)^2+4(v_u-1/2)^2+5/2 } ~, 
\end{equation}
where $a_q = 1 - v_q$ is used and the quark masses have been neglected.
Among the various decay modes, the $Z'\to b\bar{b}$ decay channel with b-tagging can be the most important one for discovering the $Z'$ boson at colliders, especially in the small mass regime $m_{Z'} \leq 2m_t$. 
Fig.~\ref{Br_contour} shows the contour plot of Br$(Z' \to b \bar b)$ on the $v_u$-$v_d$ plane.
The branching ratio increases (decreases) as $v_d$ ($v_u$) deviates from $1/2$.

\section{Constraints \label{Sec:const}}

\begin{figure}[t]
\begin{center}
\includegraphics[width=80mm]{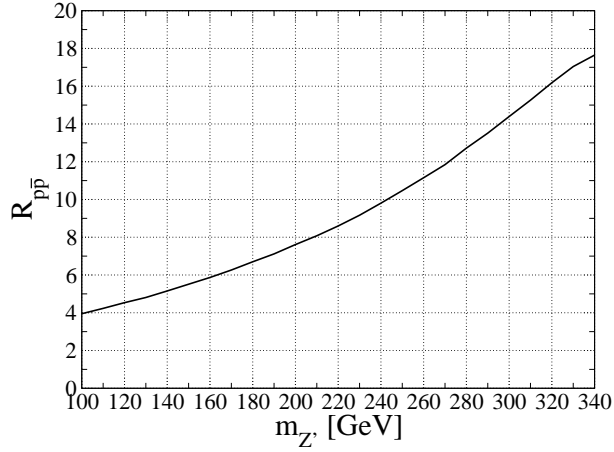}
\end{center}
\caption{${\cal R}_{p\bar{p}}$ as a function of $m_{Z'}$ at the CM energy of 630 GeV using the ${\tt CTEQ6L}$ PDF's. }
\label{fig:R}
\end{figure}

We first consider the constraint on the coupling constants of the $Z'$ boson to quarks from the UA2 experiment. 
The UA2 experiment had searched for a $Z'$ boson via the $p\bar{p}\to Z^{\prime(*) } \to jj$ process at the center-of-mass (CM) energy of 630 GeV. 
The analysis was done in the mass range from 100 GeV to 300 GeV~\cite{UA2}.   
The cross section for the hard process $q\bar{q}\to Z^{\prime(*)} \to q'\bar{q}'$ is given by 
\begin{align}
\hat{\sigma}_{q\bar{q} \to q'\bar{q}'}(\hat{s}) = \frac{g_{Z'}^4}{12\pi \hat{s}}
\frac{ (v_q^2 + a_q^2)(v_{q'}^2 + a_{q'}^2)}{(1-m_{Z'}^2/\hat{s})^2+m_{Z'}^2\Gamma_{Z'}^2/\hat{s}^2}, 
\end{align}
where $\sqrt{\hat{s}}$ is the CM energy of the partons $q$ and $\bar q$.
As can be inferred from Eq.~(\ref{eq:partialwidth}), $\Gamma_{Z'} / m_{Z'} \sim g_{Z'}^2/(4\pi)$ whose value for our benchmark points defined below is of ${\cal O}(0.01)$.  Therefore, the narrow width approximation can be employed to simplify the cross section as
\begin{align}
\hat{\sigma}_{q\bar{q} \to q'\bar{q}'}(\hat{s})& \simeq \frac{g_{Z'}^4}{12}
\frac{\hat{s} }{m_{Z'}^{}\Gamma_{Z'}}(v_q^2 + a_q^2)(v_{q'}^2 + a_{q'}^2)
\delta(\hat{s}-m_{Z'}^2)
\notag\\
&= \frac{\pi g_{Z'}^2}{3}
(v_q^2 + a_q^2)\times\text{Br}(Z'\to q'\bar{q}')
\delta(\hat{s}-m_{Z'}^2). 
\end{align}
The cross section for the $p\bar{p}\to Z^{\prime *}\to q'\bar{q}'$ process is obtained by convoluting the above expression with the partonic luminosity functions $d\mathcal{L}_{q\bar{q}}/d\tau(\tau,\mu_F)$ for the $q\bar{q}$ initial state as 
\begin{align}
\sigma(p\bar{p}\to Z^{\prime(*)} \to q'\bar{q}')=\sum_{q,\bar{q}}\int_{0}^1 d\tau \frac{d\mathcal{L}_{q\bar{q}}}{d\tau}(\tau,\mu_F)\hat{\sigma}_{q\bar{q}\to q'\bar{q}'}(\hat{s}=\tau s), 
\end{align}
where $\mu_F$ is the factorization scale, and $\sqrt{s}=630$ GeV. 
The luminosity function is given by
\begin{align}
\mathcal{L}_{q\bar{q}}(\tau,\mu_F) = \int_\tau^1 \frac{dx}{x}f_q(x,\mu_F)f_{\bar{q}}(\tau/x,\mu_F),
\end{align}
with $f_q$ and $f_{\bar{q}}$ being the parton distribution functions (PDF's) of $q$ and $\bar{q}$, respectively.  
Again, in the narrow width approximation, the cross section becomes 
\begin{align}
&\sigma(p\bar{p}\to Z^\prime \to q'\bar{q}') \simeq 
\sum_{q,\bar{q}} \mathcal{L}_{q\bar{q}}(m_{Z'}^2/s,\mu_F)\hat{\sigma}_{q\bar{q}\to q'\bar{q}'}(\hat{s} =  m_{Z'}^2)\notag\\
& \simeq \frac{2\pi g_{Z'}^2}{3} \mathcal{L}_{d\bar{d}}
\left[{\cal R}_{p\bar{p}}\left(v_u-\frac{1}{2}\right)^2+\left(v_d-\frac{1}{2}\right)^2+\frac{1}{4}({\cal R}_{p\bar{p}}+1)\right]\text{Br}(Z'\to q'\bar{q}'), 
\label{XS_Rexp}
\end{align}
where ${\cal R}_{p\bar{p}} \equiv \mathcal{L}_{u\bar{u}}/\mathcal{L}_{d\bar{d}}$ denotes the ratio of luminosities of the $u\bar{u}$ and $d\bar{d}$ initial states. 
It is seen that the cross section is proportional to the elliptical expression of $v_u$ and $v_d$ that has dependence on ${\cal R}_{p\bar{p}}$. 
In Fig.~\ref{fig:R}, we show ${\cal R}_{p\bar{p}}$ as a function of $m_{Z'}$ in the mass of interest to us.  Here $\mu_F=m_{Z'}$, $\sqrt{s}=630$, and the ${\tt CTEQ6L}$ PDF's are used.

\begin{figure}[t]
\begin{center}
\includegraphics[width=75mm]{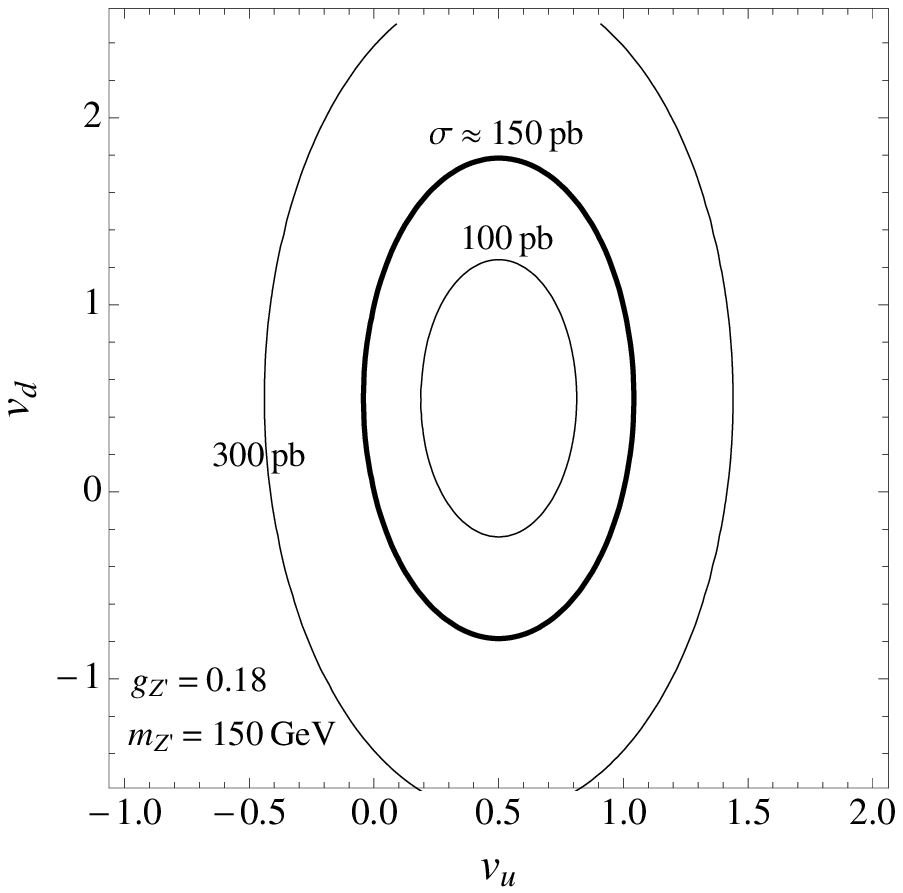} \hspace{5mm}
\hspace{0.5cm}
\includegraphics[width=75mm]{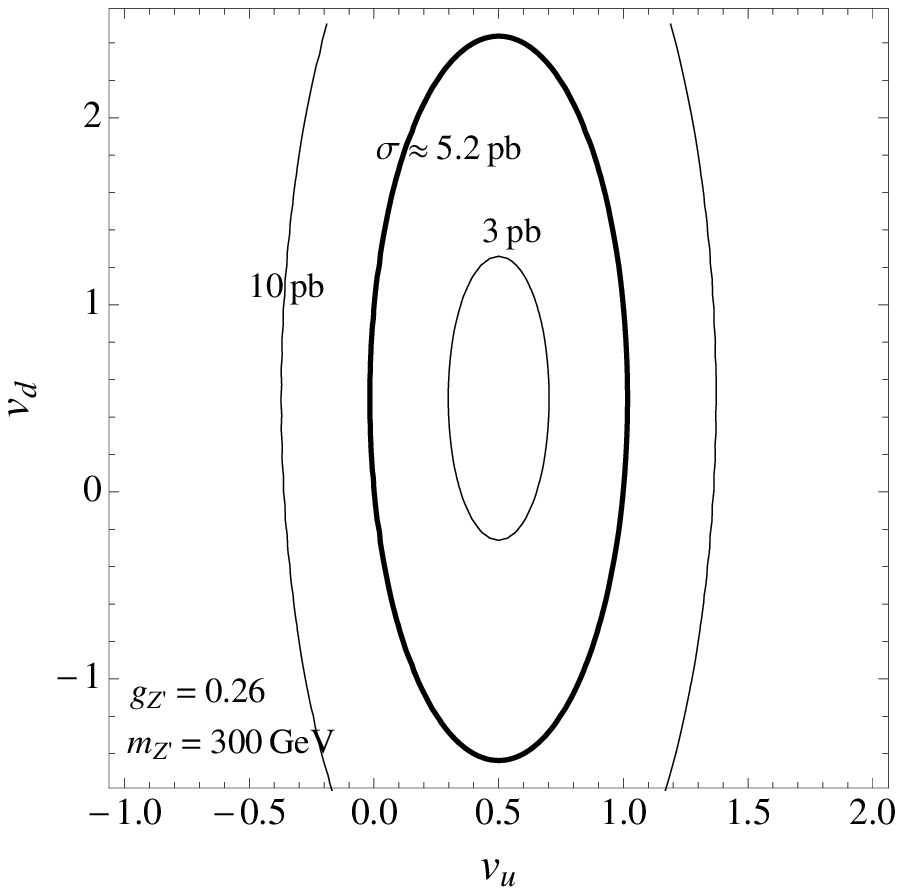}
\end{center}
\caption{Contours of the $p \bar{p} \to Z' \to jj$ cross section with the CM energy of 630 GeV on the $v_u$-$v_d$ plane, where $g_{Z'} = 0.18$ ($0.26$) and $m_{Z'} =150~(300)$ GeV are applied to the left (right) plot. 
The thick black curves correspond to the maximum cross section allowed by the UA2 experiment.  }
\label{cross_contour}
\end{figure}

According to Eq.~(\ref{XS_Rexp}), the maximum of $v_u~(v_d)$ for given values of $g_{Z'}$ and the cross section is obtained when we fix $v_d~(v_u) = 1/2$. 
In particular, the same cross section for ($v_u$,$v_d$)=(1,1) is obtained by taking ($v_u$,$v_d$)=(1/2,$v_d^{\text{max}}$) and ($v_u$,$v_d$)=($v_u^{\text{max}}$,1/2) for the same value of $g_{Z'}^{}$, where 
\begin{align}
v_{u}^{\rm max} &=\sqrt{\frac{{\cal R}_{p\bar{p}}+1}{2 {\cal R}_{p\bar{p}}}}+\frac{1}{2},~~(\text{for}~~v_d=1/2), \\ 
v_{d}^{\rm max} &=\frac{1}{2} (1+\sqrt{{\cal R}_{p\bar{p}}+1}),~~  (\text{for}~~v_u=1/2).   \label{vdmax}
\end{align}

In Fig.~\ref{cross_contour}, we show the contours of the $p \bar{p} \to Z' \to jj$ cross section for $m_{Z'} = 150$ GeV and $g_{Z'} = 0.18$ 
($m_{Z'} = 300$ GeV and $g_{Z'} = 0.26$) in the left (right) plot.  
These values of $g_{Z'}$ are obtained for the choice of ($v_u$,$v_d$)=(1,1).
The thick contour in the left (right) plot corresponds to $\sigma \simeq 150$ (5.2) pb, the upper limit set by the UA2 experiment~\cite{UA2}.
From the thick contours, we obtain the values of $v_d^{\text{max}}$ to be about 1.8 and 2.5 in the left and right plots, respectively, 
as can also be obtained by using Eq.~(\ref{vdmax}) with ${\cal R}_{p\bar{p}}=5.4$ and $14.3$ given in Fig.~\ref{fig:R}.

\begin{table}[t]
\begin{center}
\begin{tabular}{l|ccc|ccccccc} \hline \hline
    &  ~~$v_u$~~ & ~~$v_d$~~ & ~~$g_{Z'}^{\rm max}$~~ & ~~$a_u$~~ & ~~$a_d$~~ \\ \hline
BP1 & 1 & 1 & $g_{Z'}^{\rm max(BP1)}$ & 0 & 0 \\ \hline 
BP2 & 1/2 & $v_d^{\text{max}}$ & $g_{Z'}^{\rm max(BP2)}$ & $1/2$ & $1-v_d^{\text{max}}$  \\ \hline 
BP3 & 1/2 & 1/2 & $g_{Z'}^{\rm max(BP3)}$ & $1/2$ & $1/2$  \\ \hline 
BP4 & 0 & 1/2 & $g_{Z'}^{\rm max(BP4)}$ & 0 & $-1/2$  \\ \hline\hline 
\end{tabular}
\caption{
Four benchmark points defined by the values of their vector couplings, $v_u$ and $v_d$, and the maximum $g_{Z'}^{\rm max}$ allowed by the UA2 experiment. 
BP1, BP2 and BP3 have $\bar{\cal Q}(Q_L)=1$, while BP4 has $\bar{\cal Q}(Q_L)=0$.  The value of $g_{Z'}^{\rm max}$ is a function of $m_{Z'}$ shown in Fig.~\ref{UA2Limit}.  Also shown are the corresponding axial-vector couplings $a_u$ and $a_d$ for quick reference.
}
 \label{BPs}
\end{center}
\end{table}

\begin{figure}[t]
\begin{center}
\includegraphics[width=80mm]{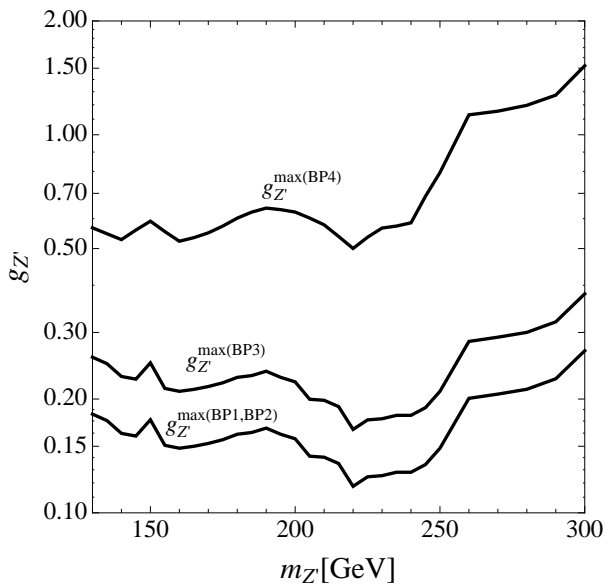}
\end{center}
\caption{The upper limit of the coupling of $Z'$ for several benchmark points which are obtained from the $p \bar p \to j j$ cross section measurement of the UA2 experiment.}
\label{UA2Limit}
\end{figure}

In the following, we estimate the upper limit on the gauge coupling $g_{Z'}^{}$ using the UA2 data for several benchmark points (BP's) of ($v_u$,$v_d$) defined in Table~\ref{BPs}.  
BP1 corresponds to the one-Higgs doublet case.  
BP2 has the maximum value for $v_d$ using $g_{Z'}^{\rm max(BP1)}$, the upper limit of $g_{Z'}^{}$ for BP1.  
BP3 corresponds to the case with purely left-handed $Z'$ couplings; {\it i.e.}, $g_{R}^{q}$ given in Eq.~(\ref{lrcoupling}) vanishes. 
BP4 is the case with nonzero $Z'$ couplings only for the right-handed down-type quarks; {\it i.e.}, $\bar{\cal Q}(Q_L)=\bar{\cal Q}(u_R)=0$.

We compute the cross section of the $p \bar p \rightarrow Z' \rightarrow j j$ process
at the CM energy of 630 GeV using {\tt MADGRAPH/MADEVENT\,5}~\cite{Ref:MG}
and our model files, and find the maximum coupling $g_{Z'}^{\rm max}$ for each benchmark point that saturates the cross section upper bound at 90\% confidence level (CL) shown in Fig.~2 of Ref.~\cite{UA2}.
In this calculation, we take the $Z'$ width calculated by {\tt CalcHEP\,3.6.15}~\cite{CalcHEP}.
We also apply a global K-factor $K=1.3$ for the cross section~\cite{Barger:1987nn}. 
Fig.~\ref{UA2Limit} shows the upper limit of the gauge coupling for each BP, 
where the limit for BP2 is taken to be the same as that for BP1, and that for BP4 is consistent with the constraint given in Fig.~1 of Ref.~\cite{Buckley}.
We will use these BP's and the corresponding constraints in the following studies of collider phenomenology.

In the case of $m_{Z'}>2m_t$, the $s$-channel $pp\to Z' \to t\bar{t}$ process is most useful for the $Z'$ search at the LHC. 
The CMS group reported the search for production of heavy resonances decaying into $t \bar{t}$ pairs
using the data of integrated luminosity of 19.6 fb$^{-1}$ at 8 TeV~\cite{CMStt}. 
Nonobservation of an excess in this process provided an upper limit on the cross section ($pp \to Z'$) times the branching fraction of $Z'\to t \bar{t}$ at 95\% CL as a function of the invariant mass $M_{t \bar{t}}$ of the $t \bar{t}$ pair.
Comparing this limit with the cross sections of $pp\to Z'\to t\bar{t}$ computed for our scenarios, 
one can also obtain the constraint on $g_{Z'}$ as a function of $m_{Z'}$. 
In our calculation, we apply a global K-factor of $K=1.4$ taken from Ref.~\cite{Gao:2010bb}. 
This bound in the $m_{Z'}>2m_t$ regime will be imposed on the study of each BP in the next section. 

\section{Collider Phenomenology}

\begin{figure}[t]
\begin{center}
\includegraphics[width=80mm]{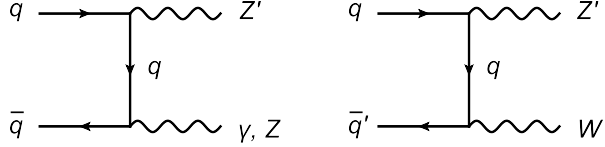}
\end{center}
\caption{Feynman diagrams for the gauge boson associated production process of the $Z'$ boson at the LHC. }
\label{fig:diagram}
\end{figure}

\subsection{Gauge boson associated production of $Z'$ \label{sec:Z'V}}

Consider the $t$-channel production of the $Z'$ boson associated with a gauge boson $V~(=\gamma,~Z$ or $W^\pm)$ at the LHC, {\it i.e.}, the $pp \to Z' V$ processes.
The dominant contributions of the $Z'V$ production are given by quark initial states $q \bar q ~(q \bar q') \to Z' V$ as shown in Fig.~\ref{fig:diagram}.  For the $Z' \gamma,~Z'Z$ final states, additional contributions come from the gluon fusion processes through quark box diagrams.  Their cross sections are only a few percent of the dominant processes at the LHC~\cite{Campbell:2011bn}. We thus neglect the gluon fusion contributions in the following analysis.
The $pp\to Z'V$ cross section is given by 
\begin{align}
\sigma(pp\to Z'V) \propto &~   
{\cal R}_{pp}(v_{V,u}^{\text{SM}}-a_{V,u}^{\text{SM}})^2\left(v_u-\frac{1}{2}\right)^2
+ (v_{V,d}^{\text{SM}}-a_{V,d}^{\text{SM}})^2\left(v_d-\frac{1}{2}\right)^2\notag\\
& +\frac{1}{4}{\cal R}_{pp} (v_{V,u}^{\text{SM}}+a_{V,u}^{\text{SM}})^2 + \frac{1}{4}(v_{V,d}^{\text{SM}}+a_{V,d}^{\text{SM}})^2, \label{cross_ZpV}
\end{align}
where the SM vector and axial-vector couplings
\begin{align}
\left(v_{V,q}^{\text{SM}},a_{V,q}^{\text{SM}}\right)& = \left(Q_q,0\right),~~~~~~~~~~~~~~~~~~~~~~~~~~\text{for}~~V=\gamma, \notag\\
\left(v_{V,q}^{\text{SM}},a_{V,q}^{\text{SM}}\right)& = \left(I_q/2-Q_q\sin^2\theta_W ,I_q/2\right),~~\text{for}~~V=Z, \notag\\ 
\left(v_{V,q}^{\text{SM}},a_{V,q}^{\text{SM}}\right)& = \left(1,1\right),~~~~~~~~~~~~~~~~~~~~~~~~~~~~\text{for}~~V=W,   \label{vsm}
\end{align}
with $Q_q$ and $I_q$ being respectively the electric charge and the third isospin component of the quark $q$.  In Eq.~(\ref{cross_ZpV}), ${\cal R}_{pp}$ is the ratio of the luminosity functions for the $pp$ collision, analogous to ${\cal R}_{p\bar p}$ in Eq.~(\ref{XS_Rexp}). 
For the collision energy of 8 TeV and $m_{Z'}^{}=150$ (300) GeV, ${\cal R}_{pp}$ is found to be about 1.5 (1.6) using the {\tt CTEQ6L} PDF's. 

\begin{figure}[t]
\begin{center}
\includegraphics[width=75mm]{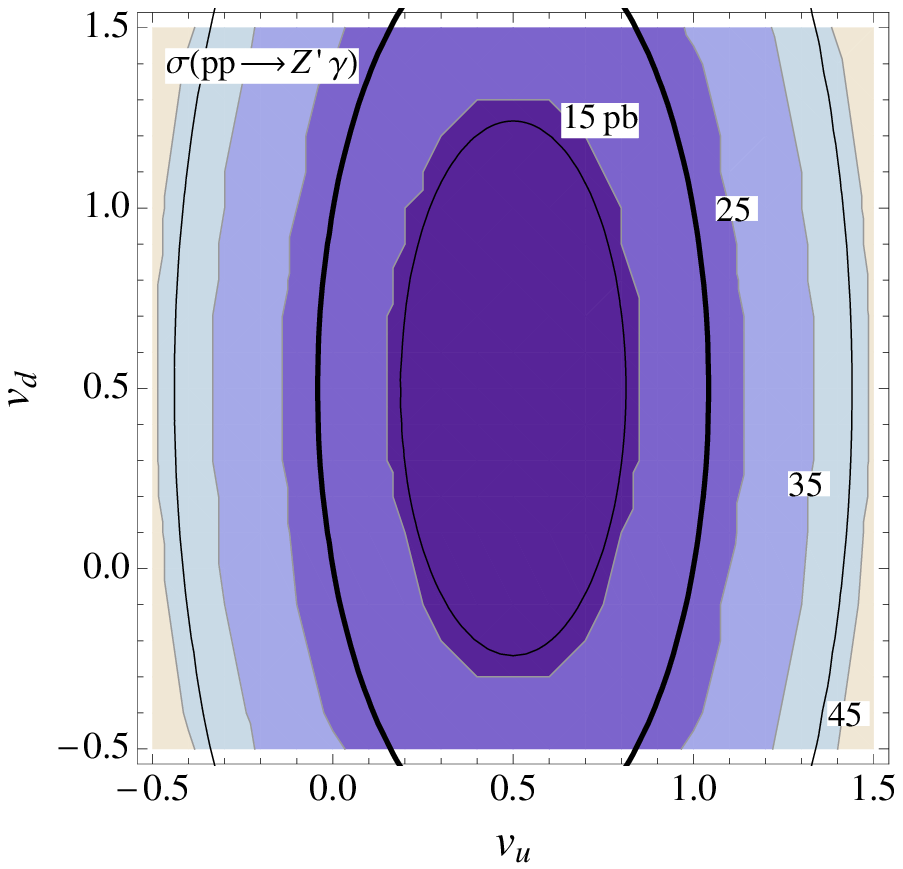} \hspace{5mm}
\hspace{0.5cm}
\includegraphics[width=75mm]{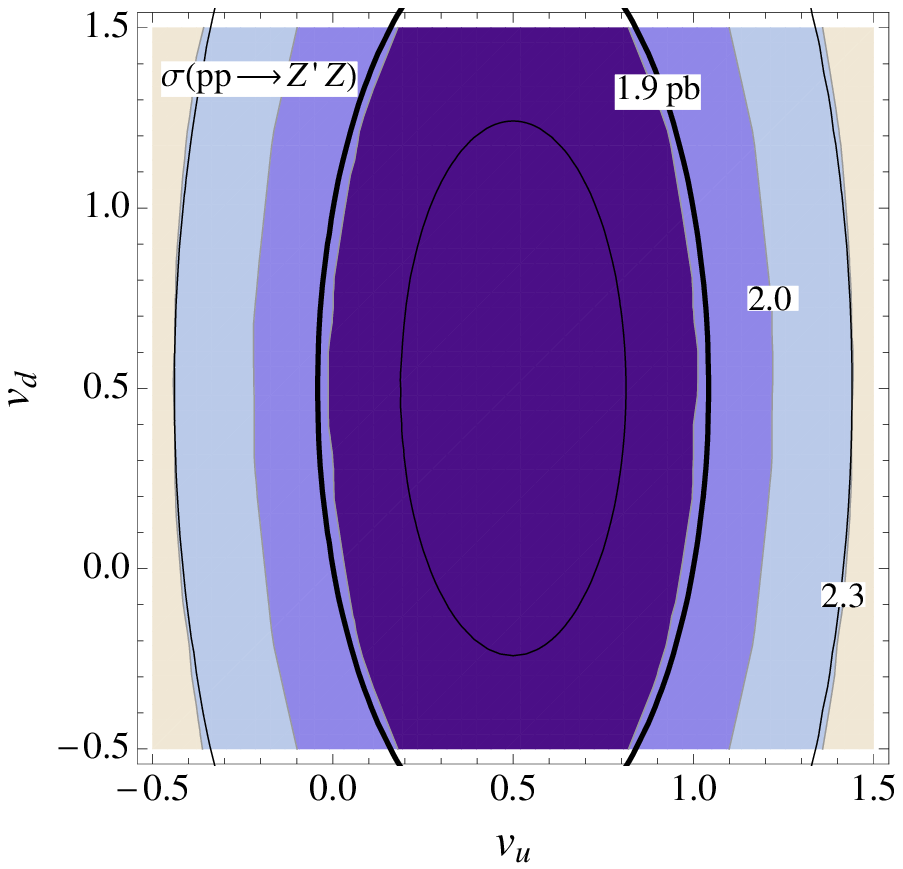}
\end{center}
\caption{Contour plots for the cross section of the $pp \to Z' \gamma$ (left) and $pp \to Z' Z$ (right) processes for the 8-TeV LHC on the $v_u$-$v_d$ plane, where $g_{Z'} = 0.18$ and $m_{Z'} =150$ GeV are applied.  The black curves correspond to those in Fig.~\ref{cross_contour}.  }
\label{cross_contour_ZpV}
\end{figure}

Fig.~\ref{cross_contour_ZpV} shows the contour plots of the cross section for the $Z'\gamma$ (left) 
and $Z'Z$ (right) final states on the ($v_u$,$v_d$) plane, where we take $g_{Z'} = 0.18$ and $m_{Z'} = 150$ GeV as in Fig.~\ref{cross_contour}.
Contours of the $p \bar p \to Z' \to jj$ cross sections in Fig.~\ref{cross_contour} are also shown in the plots for comparison.
The $pp\to Z'W$ cross section is constant, about 5.5 pb in the case of $g_{Z'} = 0.18$ and $m_{Z'} = 150$ GeV, on the ($v_u$,$v_d$) plane as shown by Eqs.~(\ref{cross_ZpV}) and (\ref{vsm}). 
It is observed that the cross sections of the $pp \to Z' \gamma$ and $pp \to Z' Z$ processes 
have similar dependence on $(v_u, v_d)$ as that of the $p \bar p \to Z' \to jj$ process.
This can be readily understood as follows: the dependence on $(v_u, v_d)$ is determined by the elliptical expression given in Eq.~(\ref{cross_ZpV}) similar to the cross section of $p \bar p \to Z' \to jj$. 
The shape of ellipse is determined by $(Q_u/Q_d)^2{\cal R}_{pp}$ according to Eqs.~(\ref{cross_ZpV}) and (\ref{vsm}). 
For the case of $m_{Z'}^{}=150$ GeV, this factor is about 6, close to ${\cal R}_{p\bar{p}}$. 

As mentioned in Section~\ref{sec:ZprDecay}, the $Z'\to b\bar{b}$ decay is the most promising mode to search for $Z'$ in the low-mass regime as one can use b-tagging to reduce backgrounds.  
We consider the signal events 
\begin{align}
& p p \to Z' \gamma \to b \bar b \gamma\to j_b j_b\gamma, \label{g_sig}\\
& p p \to Z' Z \to b \bar b \ell^+ \ell^- \to j_b j_b\ell^+ \ell^-, \label{z_sig} \\
& p p \to Z' W^\pm \to b \bar b \ell^\pm E_T\hspace{-4.5mm}/ \hspace{2mm} \to j_b j_b \ell^\pm E_T\hspace{-4.5mm}/ \hspace{2mm}, \label{w_sig}
\end{align}
where $\ell =e$ or $\mu$, $E_T\hspace{-4.5mm}/ \hspace{2mm}$ is the missing transverse energy, and $j_b$ is a tagged b-jet.
For these production processes, we also take into account the enhancement from QCD corrections characterised by the K-factor $K=1.3$~\cite{Ohnemus:1992jn, Baur:1994aj} in our analysis.
The SM backgrounds corresponding to each of above signals are
\begin{align}
& p p \to b \bar b \gamma \to j_b j_b\gamma, \quad \quad \quad \quad p p \to j j \gamma\to j_b j_b \gamma,  \label{g_bg}\\
& p p \to b \bar b \ell^+ \ell^- \to j_b  j_b \ell^+ \ell^- , \quad p p \to j j \ell^+ \ell^- \to j_b j_b \ell^+ \ell^-, \label{z_bg}\\
& p p \to b \bar b \ell^\pm E_T\hspace{-4.5mm}/ \hspace{2mm}\to j_b j_b \ell^\pm E_T\hspace{-4.5mm}/ \hspace{2mm}, 
\quad p p \to jj \ell^\pm E_T\hspace{-4.5mm}/ \hspace{2mm}\to j_b j_b \ell^\pm E_T\hspace{-4.5mm}/ \hspace{2mm},  \label{w_bg}
\end{align}
where $j$ is a jet coming from a gluon or a non-bottom quark. 
Since the backgrounds involve various processes with different K-factors, some of which have not been evaluated yet, we take $K= 1.2$ and 1.4 to estimate possible uncertainties.
The signal and background events are both generated using {\tt MADGRAPH/MADEVENT\,5}, and passed to {\tt PYTHIA\,6}~\cite{Ref:Pythia} via the {\tt PYTHIA-PGS} package to include initial-state radiation, final-state radiation and hadronization effects.
We note in passing that in {\tt MADGRAPH/MADEVENT\,5} the factorization and the renormalization scales are set as $\sum_{i=1}^2 (M_i^2 + p_{T i}^2)/2$ for the two final-state particles.
The detector-level simulation is carried out using {\tt PGS\,4}~\cite{Ref:PGS}, which performs the b-tagging with an efficiency about $0.5$ for high-energy jets in the central region defined by the jet rapidity limit $|\eta(j)| < 2.0$. 
The number of signal events is reduced to $\sim 10 \%$ due to double b-tagging and the rapidity cut.

In addition, since the b-jets are energetic and boosted along the direction of $Z'$, we thus impose the following kinematic cuts for the transverse momentum of each b-jet, $p_T(j_b)$, and the rapidity difference between the two b-jets, $\Delta \eta_{j_b j_b}$:
\begin{align}
\label{jetCuts}
& p_T( j_b) > 40~{\rm GeV}, \quad
|\Delta \eta_{j_bj_b}| <2.0, 
\end{align} 
where the lower and upper limit on $p_T(j_b)$ and $|\Delta \eta_{j_bj_b}|$ are chosen by optimizing the cut efficiency at $m_{Z'} = 150$ GeV.  For the $j_b j_b \gamma$ events, we further eliminate soft photons by applying the following $p_T$ cut:
\begin{align}
p_T(\gamma)  > 10~{\rm GeV}~~\text{for}~~ j_b j_b\gamma ~,  \label{PhotonCut}
\end{align}
where we have not chosen a larger lower limit for $p_T(\gamma)$ as the photon energy in $Z' \gamma$ production process tends to be small.
We also take the following cuts for $j_b j_b\ell^+\ell^-$ and $j_b j_b \ell^\pm E_T\hspace{-4.3mm}/ \hspace{2mm}$ events:
\begin{align}
p_T({\rm \ell}) &> 25~{\rm GeV}~~\text{for}~~ j_b j_b\ell^+\ell^-~~\text{and}~~j_b j_b\ell^\pm E_T\hspace{-4.3mm}/ \hspace{2mm} ~, \notag\\ 
E_T\hspace{-4.5mm}/ \hspace{2mm} &> 25~{\rm GeV}~~\text{for}~~j_b j_b\ell^\pm E_T\hspace{-4.5mm}/ \hspace{2mm} ~,  \label{leptonCut}
\end{align}
where the lower limit for $p_T(\ell)$ is taken from the $W^+ b \bar b$ search at CMS~\cite{Chatrchyan:2013uza} and the same value is used for $E_T\hspace{-4.5mm}/ \hspace{2mm}$.
Finally, we also make a cut on the invariant mass of the two b-jets $M_{j_bj_b}$; 
\begin{equation}
\label{invariantMass}
m_{Z'} (1-0.2) < M_{j_bj_b} < m_{Z'} + 10 \ {\rm GeV} ~.
\end{equation}
Here we have chosen asymmetric limits, as also used in our previous paper~\cite{Chiang:2014yva}, because the shape of $b \bar b$ invariant mass distribution is not symmetric around $m_{Z'}$.

\subsection{One-Higgs doublet case}

\begin{figure}[t]
\begin{center}
\includegraphics[width=70mm]{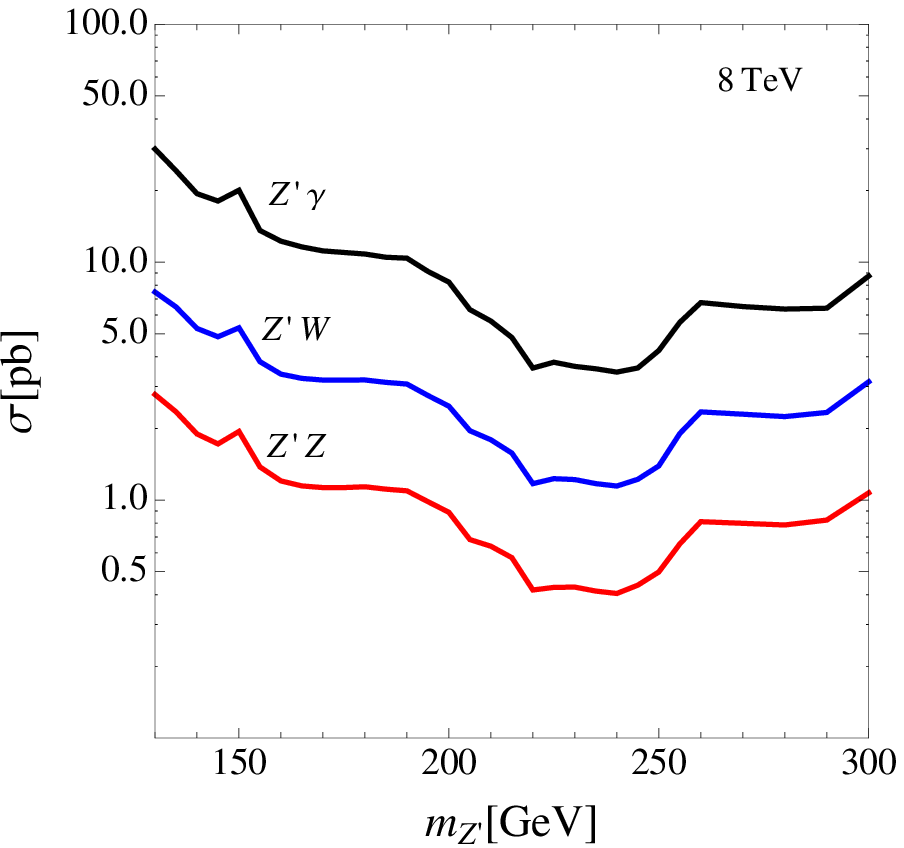}\hspace{8mm}
\hspace{0.5cm}
\includegraphics[width=70mm]{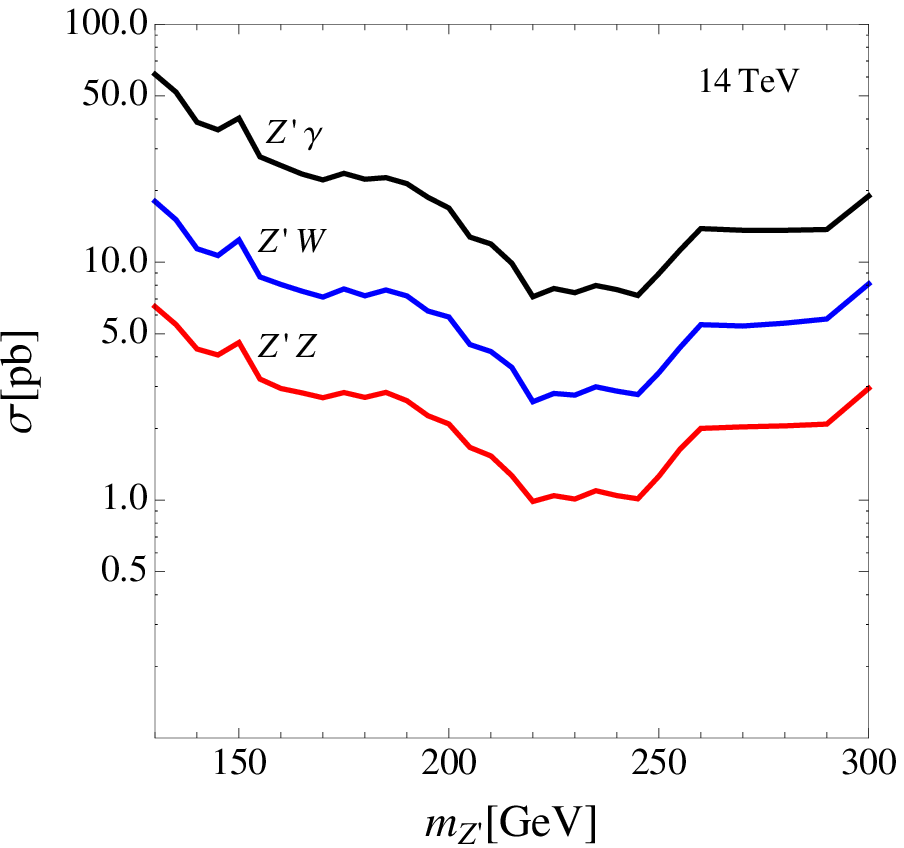}
\end{center}
\caption{Cross sections of the $pp\to Z'V$ ($V=\gamma$, $W$ and $Z$) processes for the one-Higgs doublet case (BP1) at the 8-TeV (left) and 14-TeV (right) LHC. }
\label{xs1h}
\end{figure}

We first apply the above analysis to the one-Higgs doublet case (or BP1), where $v_u=v_d=1$. Here we remind the reader that each BP by definition takes the maximum gauge coupling, $g_{Z'}^{\rm max}$, that saturates the UA2 bound.  Therefore, the cross sections given in the following numerical calculations are their maxima derived from the curves given in Fig.~\ref{UA2Limit}.
In Fig.~\ref{xs1h}, the cross sections of the $pp\to Z'V$ processes at the 8-TeV (left plot) and 14-TeV (right plot) LHC are given as functions of $m_{Z'}$,
all computed with {\tt CalcHEP}. 
It is seen that the $pp\to Z'\gamma$ process gives a $\sim 4$ and 10 times larger cross section than the $pp\to Z'W$ and $pp\to Z'Z$ processes, respectively. 

\begin{table}[t]
\begin{center}
\footnotesize
\begin{tabular}{l|ccc|ccc|ccc} \hline \hline
Events & \multicolumn{3}{c|}{$j_bj_b \gamma$} & \multicolumn{3}{c|}{$j_bj_b \ell^+ \ell^-$} & \multicolumn{3}{c}{$j_bj_b \ell^\pm E_T$}   \\ \hline
            & $\sigma_S$~[pb] & $\sigma_B^{\rm LO}$~[pb] & $\mathcal{S}$ & $\sigma_S$~[pb] & $\sigma_B^{\rm LO}$~[pb] & $\mathcal{S}$ & $\sigma_S$~[pb] & $\sigma_B^{\rm LO}$~[pb] & $\mathcal{S}$ \\
b-tagging & 1.6$\times 10^{-1}$ & 18. & 4.9 (4.6) & 1.1$\times 10^{-3}$ & 1.1$\times 10^{-1}$ & 0.43 (0.40) & 1.3$\times 10^{-2}$ & 5.9$\times 10^{-1}$ & 2.2 (2.1)   \\ 
$p_T(j_b)>40$ GeV & 1.0$\times 10^{-1}$ & 4.8 & 5.9 (5.4) & 6.5 $\times 10^{-4}$ & 2.8$\times 10^{-2}$ & 0.49 (0.45) & 7.7$\times 10^{-3}$ & 1.9$\times 10^{-1}$ & 2.2 (2.1)    \\
$|\Delta \eta_{j_b j_b}|<2.0$ & 1.0$\times 10^{-1}$ & 4.5 & 6.0 (5.5) & 6.3$\times 10^{-4}$ & 2.6$\times 10^{-2}$ & 0.50 (0.47) & 7.6$\times 10^{-3}$ & 1.8$\times 10^{-1}$ & 2.3 (2.1) \\
$p_T(\ell) > 25$ GeV & & & & 4.4$\times 10^{-4}$ & 1.1$\times 10^{-2}$ & 0.52 (0.48) & 6.3$\times 10^{-3}$ & 1.1$\times 10^{-1}$ & 2.4 (2.2) \\
$E_T\hspace{-3.5mm}/ \hspace{2mm} > 25~{\rm GeV}$ & & & & & & & 4.9$\times 10^{-3}$ & 6.3$\times 10^{-2}$ & 2.5 (2.3) \\
$M_{j_bj_b}$ cut   & 6.7$\times 10^{-2}$ & 1.6 & 6.9 (6.4) & 3.0$\times 10^{-4}$ & 3.4$\times 10^{-3}$ & 0.64 (0.59) & 3.4$\times 10^{-3}$ & 1.8$\times 10^{-2}$ & 3.1 (2.9) 
\\ \hline \hline
\end{tabular}
\caption{Cross sections of signals including the K-factor $K =1.3$ ($\sigma_S$) and background processes at leading order ($\sigma_B^{\rm LO}$) after sequentially imposing each of the selection cuts shown in the first column in the case of $m_{Z'}=150$ GeV and $\sqrt{s}=8$ TeV.  
For the significance $\mathcal{S}$, we take the integrated luminosity of 19.6 fb$^{-1}$ and apply the K-factor for the background cross sections.  Values without (within) parentheses correspond to the background K-factor of $K=1.2$ (1.4).
 \label{events}}
\end{center}
\end{table}

In Table~\ref{events}, we show the cross sections for the signals and the backgrounds given in Eqs.~(\ref{g_sig})-(\ref{w_sig}) and 
in Eqs.~(\ref{g_bg})-(\ref{w_bg}), respectively, at the collision energy of 8 TeV. 
We take $m_{Z'} = 150$ GeV as an example and apply the corresponding upper limit of $g_{Z'}$ in Fig.~\ref{UA2Limit}. 
The signal significance defined as~\cite{Ref:significance}
\begin{equation}
\mathcal{S}=\sqrt{2[(s+b)\ln (s/b)-s]}, 
\label{significance}
\end{equation}
is also given in the last column of each final state with the assumption of an integrated luminosity of 19.6 fb$^{-1}$, where $s$ and $b$ denote the numbers of signal and background events, respectively.  
The number without (within) parentheses corresponds to the backgrounds using the K-factor $K=1.2$ ($1.4$). 
From the third to last rows, we show the results after sequentially imposing the kinematic cuts in the first column, as given in Eqs.~(\ref{jetCuts}), (\ref{PhotonCut}), (\ref{leptonCut}), and  Eq.~(\ref{invariantMass}).
The $Z'\gamma$ process has the largest significance ${\cal S}$ due to its largest signal cross section among all. 
Although the $Z'Z$ process has the smallest background cross section, the signal cross section 
is also highly suppressed due to the leptonic branching fraction of $Z$.  

\begin{figure}[t]
\begin{center}
\includegraphics[width=75mm]{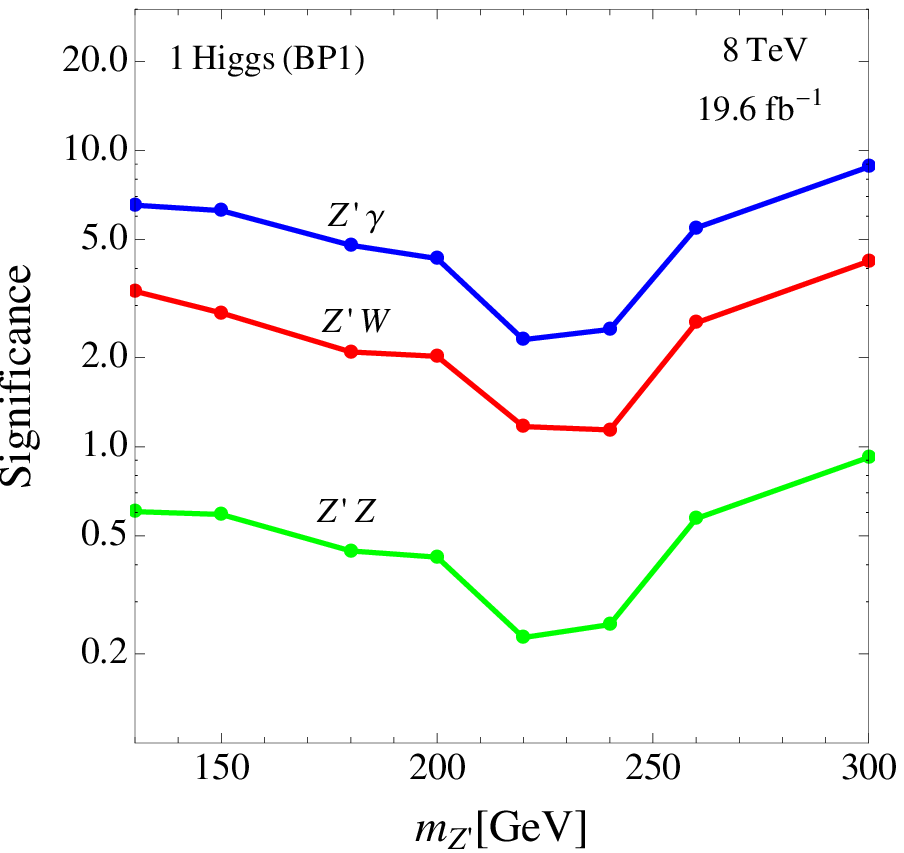} \hspace{5mm}
\hspace{0.5cm}
\includegraphics[width=75mm]{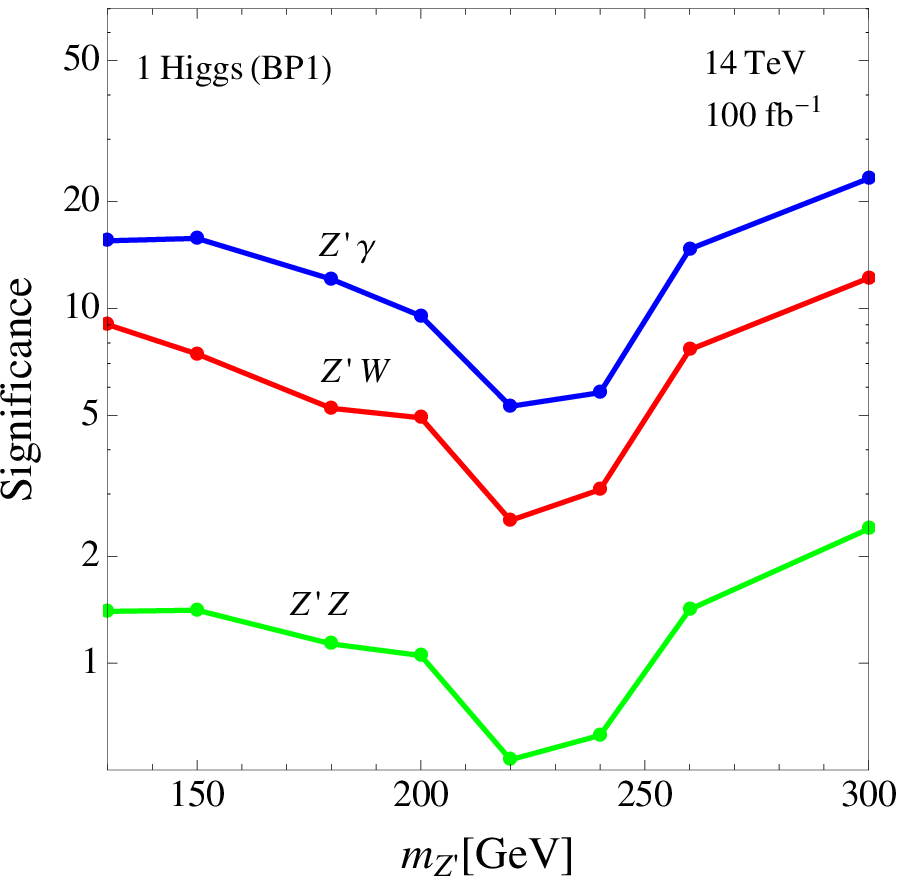}
\end{center}
\caption{ Significances for $Z' \gamma$, $Z' W$ and $Z' Z$ signals at the 8-TeV LHC with the integrated luminosity of 19.6 fb$^{-1}$ (left) and the 14-TeV LHC with the integrated luminosity of 100 fb$^{-1}$ (right).  All kinematic cuts in Eqs.~(\ref{jetCuts}), (\ref{PhotonCut}), (\ref{leptonCut}) and Eq.~(\ref{invariantMass}) have been imposed.  The K-factor of 1.4 for the backgrounds is used.
}
\label{sign_1h}
\end{figure}
%
\begin{figure}[t]
\begin{center}
\includegraphics[width=75mm]{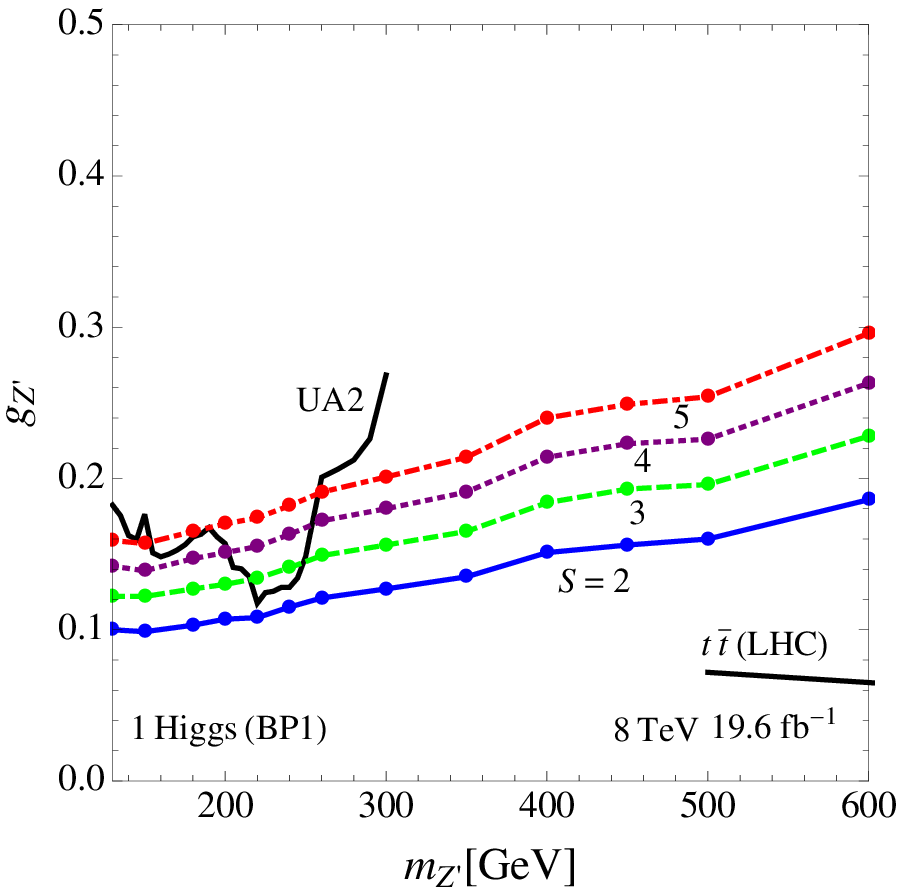} \hspace{5mm}
\hspace{0.5cm}
\includegraphics[width=75mm]{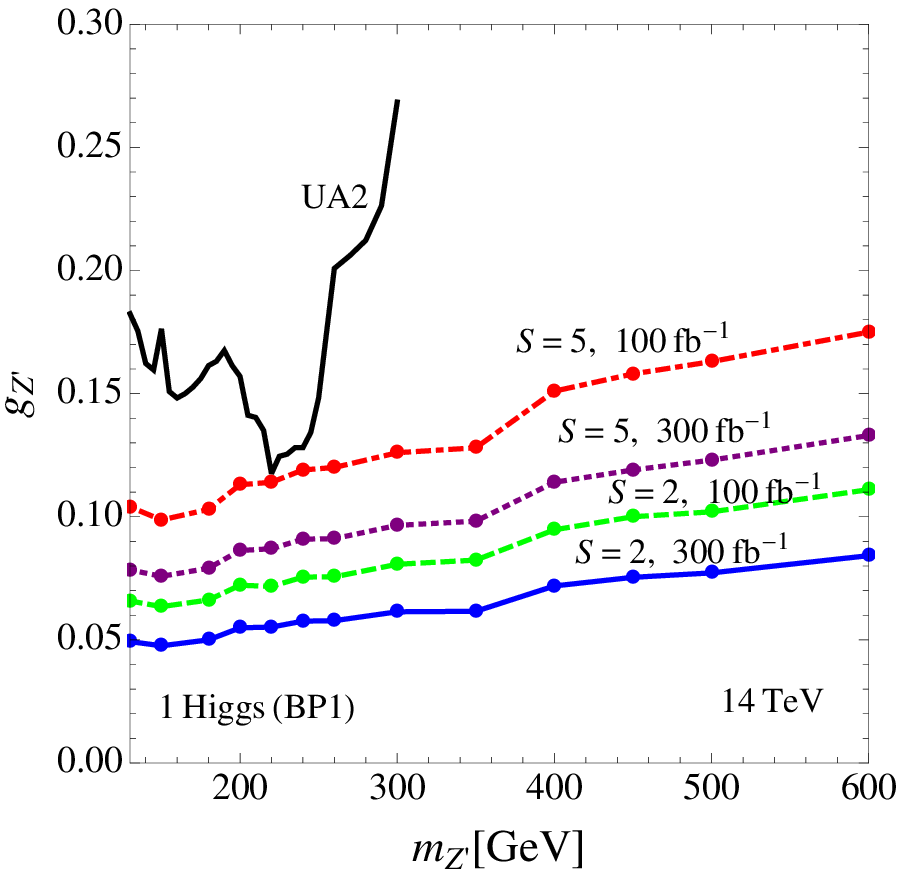}
\end{center}
\caption{The left plot shows the required gauge coupling constant to render the significance of 2, 3, 4 and 5 at the 8-TeV LHC with the integrated luminosity of $19.6$ fb$^{-1}$ for the $Z'\gamma$ process in the one-Higgs doublet case.
The right plot shows the corresponding curves for significance of 2 and 5 at the 14-TeV LHC with the integrated luminosities of 100 and 300 fb$^{-1}$.  The black curve gives the UA2 and $t \bar t$ constraints on the gauge coupling constant.}
\label{const_1h}
\end{figure}

In Fig.~\ref{sign_1h}, the signal significances for the $Z'\gamma$, $Z'Z$ and $Z'W^\pm$ processes are shown as functions of $m_{Z'}$, assuming the collision energy and the integrated luminosity respectively to be 8 TeV and 19.6 fb$^{-1}$ for the left plot and 14 TeV and 100 fb$^{-1}$ for the right plot. 
Here all the kinematic cuts in Eqs.~(\ref{jetCuts})-(\ref{invariantMass}) have been imposed and $K=1.4$ is applied to the backgrounds for a conservative estimate.  The case of $K=1.2$ for the backgrounds would have slightly better significances.  Since the UA2 upper limit of $g_{Z'}$ is used, these significances are the largest values that one can expect.
For the $Z'\gamma$ process, ${\cal S}>2$ (${\cal S}>5$) in the entire mass region of the plot ($m_{Z'}<200$ GeV and $m_{Z'} > 260$ GeV) at the 8-TeV LHC, while ${\cal S}>5$ is achieved at the 14-TeV LHC. 
The other two processes $Z'Z$ and $Z'W^\pm$ give smaller values of ${\cal S}$ as compared to that in the $Z'\gamma$ process. 
Especially for the $Z'Z$ process, ${\cal S}$ is smaller than 2 in the entire mass range considered in these figures even in the case of 14 TeV and 100 fb$^{-1}$.

As shown partly in the left plot of Fig.~\ref{const_1h} below, the $t \bar t$ data from the 8-TeV LHC has a more stringent constraint for $m_{Z'} \ge 500$ GeV.  This mode will still be the most promising search channel or impose the most stringent constraint for the same mass regime.  Between $2m_t$ and 500 GeV, the $t \bar t$ constraint is expected to be worse and may be comparable to the $Z'\gamma$ mode proposed in this work.  Therefore, we will put our focus on the region of $m_{Z'} \alt 500$ GeV in the following discussions.

The result for $Z'\gamma$ process given in Fig.~\ref{sign_1h} can be translated into the required $Z'$ coupling $g_{Z'}^{}$ to achieve a specific value of ${\cal S}$ for each value of $m_{Z'}$ using the fact that the cross section (number of events) is proportional to $g_{Z'}^2$. 
We have applied $K=1.4$ to the backgrounds in these estimates.
The left plot of Fig.~\ref{const_1h} shows 
the contours of the required $g_{Z'}^{}$ for ${\cal S}=2$, 3, 4 and 5 at the collision energy of 8 TeV and integrated luminosity of 19.6 fb$^{-1}$. 
The upper limits from the UA2 experiment is also shown in this figure. 
For the region $m_{Z'}>500$ GeV, the upper bound on $g_{Z'}^{}$ is derived from the current LHC data of the $pp\to t\bar{t}$ process~\cite{CMStt}.  
From the curve for ${\cal S}=2$, we obtain a stronger constraint on $g_{Z'}^{}$ in comparison with that from the UA2 experiment. 
Similarly, the right plot of Fig.~\ref{const_1h} shows the required $g_{Z'}^{}$ for ${\cal S}=2$ and 5 at the collision energy of 14 TeV and integrated luminosities of 100 fb$^{-1}$ and 300 fb$^{-1}$. 
The 95\% CL upper limit (${\cal S}=2$) is well below the limit from the UA2 experiment, which is about 0.06 (0.05) at $m_{Z'}=150$ GeV assuming the integrated luminosity of 100 (300) fb$^{-1}$. 
The minimal value of $g_{Z'}^{}$ for the discovery (${\cal S}=5$) is also below the UA2 limit, which is about 0.10 (0.08)
at $m_{Z'}=150$ GeV
assuming the integrated luminosity of 100 (300) fb$^{-1}$.

\subsection{$N$-Higgs doublet case}

\begin{table}[t]
\begin{center}
\begin{tabular}{l|l|llll} \hline \hline
&$m_{Z'}$ [GeV]& 150  & 200 & 250 & 300   \\ \hline
         & $\sigma(Z' \gamma)\times$Br($Z'\to b\bar{b}$) [pb]  & 4.0   & 1.6   & 8.1$\times 10^{-1}$ & 1.7 \\
BP1 &  $\sigma(Z' Z)\times$Br($Z'\to b\bar{b}$) [pb]    &   3.9$\times 10^{-1}$ \hspace{5mm} & 1.8$\times 10^{-1}$  \hspace{5mm} & 1.0$\times 10^{-1}$ \hspace{5mm} &  2.4$\times 10^{-1}$  \\ 
         &  $\sigma(Z' W)\times$Br($Z'\to b\bar{b}$) [pb]   &  1.2  & 5.2$\times 10^{-1}$  & 3.0$\times 10^{-1}$ & 6.3$\times 10^{-1}$ \\ \hline
         &  $\sigma(Z' \gamma)\times$Br($Z'\to b\bar{b}$) [pb]   & 6.1  &  2.8 & 1.7  & 3.9 \\
BP2 &  $\sigma(Z' Z)\times$Br($Z'\to b\bar{b}$) [pb]    &   6.0$\times 10^{-1}$  & 2.7$\times 10^{-1}$  &  1.6$\times 10^{-1}$ & 3.7$\times 10^{-1}$  \\ 
         &  $\sigma(Z' W)\times$Br($Z'\to b\bar{b}$) [pb]   &  1.8  & 7.8$\times 10^{-1}$ & 4.5$\times 10^{-1}$  & 9.8$\times 10^{-1}$ \\ \hline
         &  $\sigma(Z' \gamma)\times$Br($Z'\to b\bar{b}$) [pb]  & 4.0  & 1.6  & 7.0$\times 10^{-1}$ & 1.7 \\
BP3 &  $\sigma(Z' Z)\times$Br($Z'\to b\bar{b}$) [pb]    &   7.2$\times 10^{-1}$  & 3.2$\times 10^{-1}$ & 1.6$\times 10^{-1}$ & 4.1$\times 10^{-1}$ \\ 
         & $\sigma(Z' W)\times$Br($Z'\to b\bar{b}$) [pb]   &  2.4  & 1.0  & 4.9$\times 10^{-1}$ & 1.3 \\ \hline
         &  $\sigma(Z' \gamma)\times$Br($Z'\to b\bar{b}$) [pb]   & 5.7  & 2.3  & 1.1  & 2.0 \\
BP4 &  $\sigma(Z' Z)\times$Br($Z'\to b\bar{b}$) [pb]    &   3.3$\times 10^{-1}$  & 2.4$\times 10^{-1}$  & 1.4$\times 10^{-1}$ & 2.9$\times 10^{-1}$   \\ 
         &  $\sigma(Z' W)\times$Br($Z'\to b\bar{b}$) [pb]   &  0.0  & 0.0  & 0.0  & 0.0 \\ \hline\hline
\end{tabular}
\caption{Products of the $pp \to Z' V$ cross sections and the branching fraction Br$(Z' \to b \bar b)$ in units of pb for each benchmark points with the collision energy of 8 TeV assuming $m_{Z'}=$150, 200, 250 and 300 GeV.
 \label{XS_ZpV}}
\end{center}
\end{table}

For the $N$-Higgs doublet case ($N\geq 3$), we apply the analysis for the $Z'$ search discussed in Section~\ref{sec:Z'V} to the four BP's listed in Table~\ref{BPs}. 
Table \ref{XS_ZpV} shows the products of the $pp \to Z' V$ cross sections and 
the branching fraction Br$(Z' \to b \bar b)$ for several values of $m_{Z'}$ assuming the collision energy of 8 TeV. Again, the maximum coupling $g_{Z'}^{\rm max}$ in Fig.~\ref{UA2Limit} is used for each BP.
We first discuss some general properties of each BP: 
\begin{itemize}
\item BP1 is exactly same as the one Higgs doublet case discussed in the previous subsection, and is shown here for comparison.
\item BP2 has the maximum value for the $pp\to Z' \gamma \to b\bar{b}\gamma$ cross section among the four BP's as 
the branching fraction for $Z' \to b \bar b$ mode is larger than BP1 by $\sim 3/2$ owing to the larger value of $v_d$.
\item BP3 has the maximum values for the $pp\to Z' W \to b\bar{b}W$ and $pp\to Z' Z \to b\bar{b}Z$ cross sections as a consequence of the purely left-handed couplings. 
\item BP4 has nonzero couplings only for the right-handed $d$-type quarks.  Thus, the $pp \to Z'W$ cross section is identically zero.  On the other hand, the cross section of $pp\to Z' \gamma \to b\bar{b}\gamma$ is close to that in BP2 although the branching fraction for $Z' \to b \bar b$ is smaller by $\sim 2/3$.  This is because the UA2 limit on $g_{Z'}$ in BP4 is the largest.
\end{itemize}
BP2 is expected to give the largest significance for the $pp\to Z' \gamma \to b\bar{b}\gamma$ process due to its largest cross section.
The purely left-handed (right-handed) couplings in BP3 (BP4) result in distinctively different relative strengths of the production cross sections of $Z' \gamma$, $Z' W$ and $Z' Z$ from the one Higgs doublet case (BP1).  It is therefore important to compare signal cross sections between three $Z' V$ production processes to distinguish the BPs.  
In the following, we concentrate on BP2 and BP3, and estimate the signal significances and the required coupling strength.
BP4 gives a significance similar to BP2 in the case of $Z'\gamma$, and its significances for other two processes are too small to be useful.
%

\begin{figure}[t]
\begin{center}
\includegraphics[width=75mm]{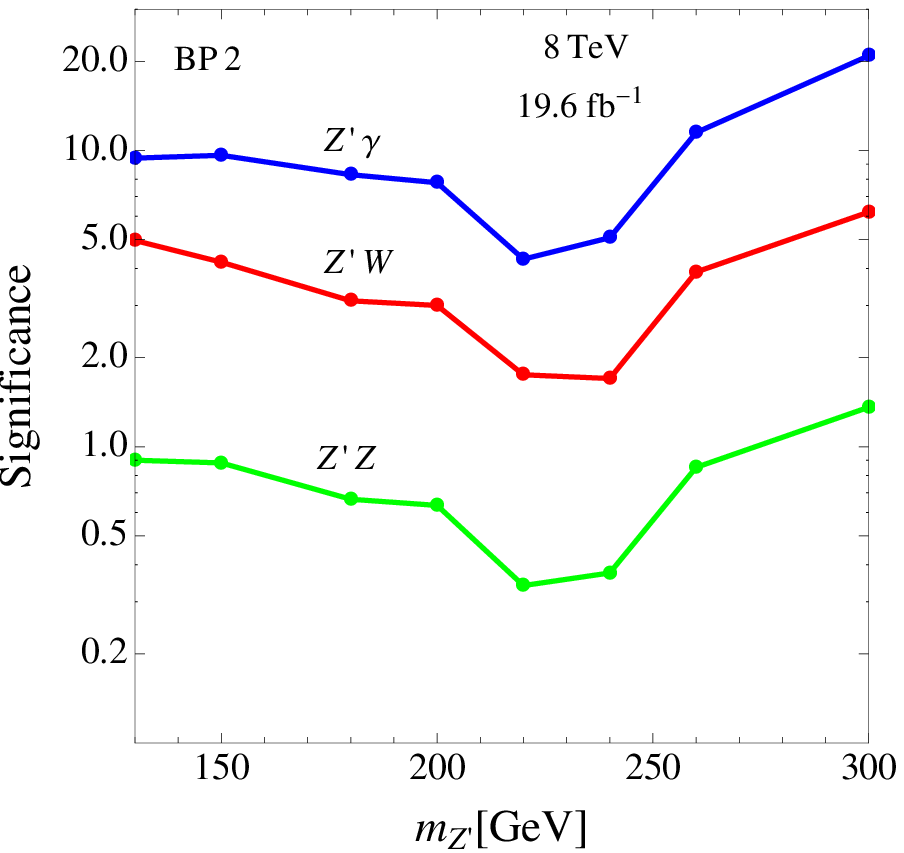} 
\hspace{0.5cm}
\includegraphics[width=75mm]{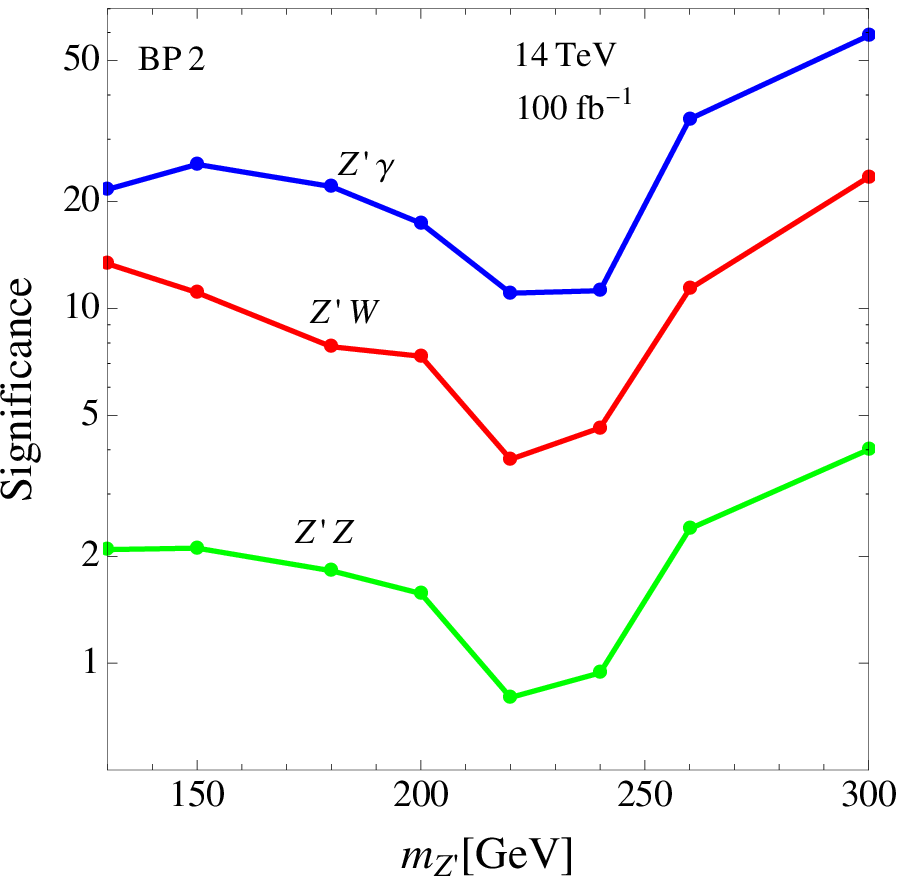}
\hspace{0.5cm}
\includegraphics[width=75mm]{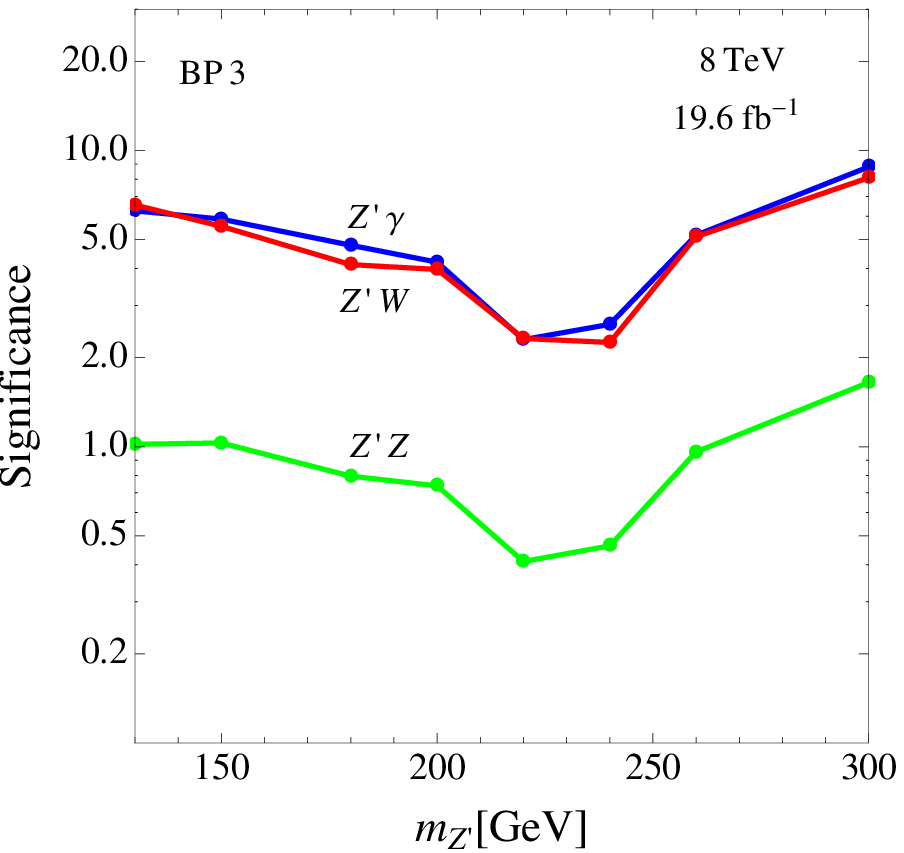} 
\hspace{0.5cm}
\includegraphics[width=75mm]{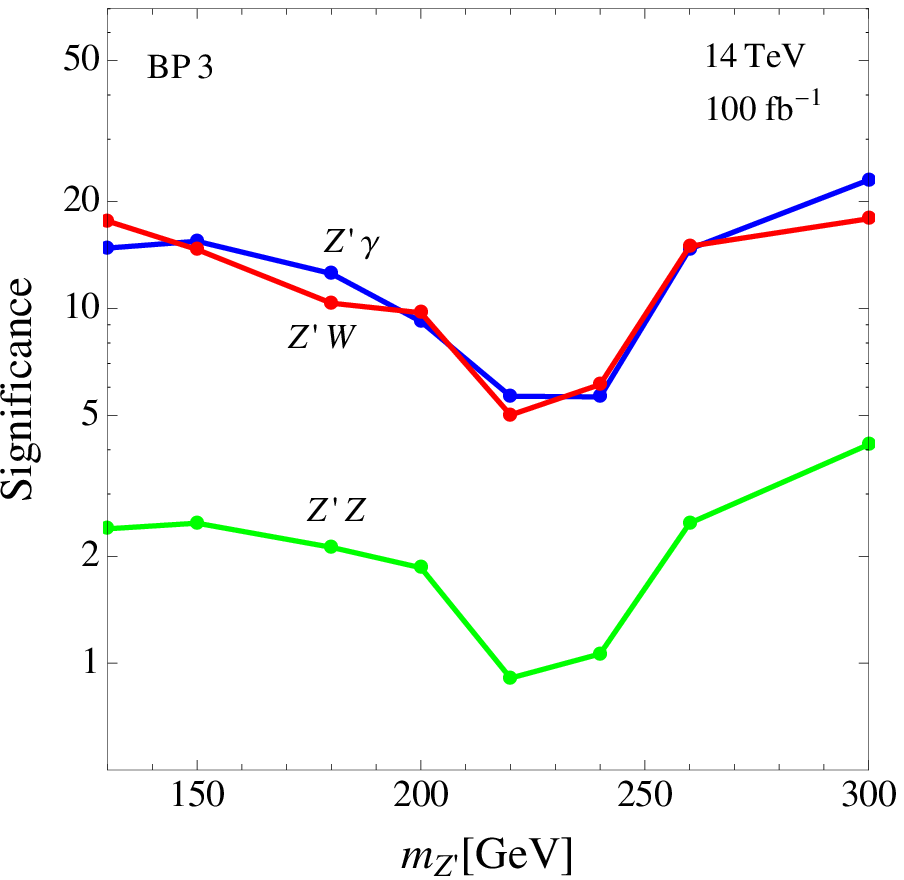}
\end{center}
\caption{Same as Fig.~\ref{sign_1h}, but for BP2 (upper plots) and BP3 (lower plots).
}
\label{Sig1}
\end{figure}

%
The upper (lower) plots of Fig.~\ref{Sig1} show the signal significances of the $Z'\gamma$, $Z'Z$ and $Z'W^\pm$ processes as functions of $m_{Z'}$ for BP2 (BP3),
assuming the collision energy and integrated luminosity of 8 TeV and 19.6 fb$^{-1}$ (left plots) and 14 TeV and 100 fb$^{-1}$ (right plots). 
Here K=1.4 is applied to the backgrounds as in the one Higgs-doublet case.
For BP2, we obtain larger significances than BP1 in Fig.~\ref{sign_1h} because of the larger branching fraction $Br(Z' \to b \bar b)$.  The hierarchy of significances is similar to that in BP1.
${\cal S}>5$ for the $Z' \gamma$ process in the entire mass range for the 8-TeV case, and the other two processes give smaller values of ${\cal S}$.
For BP3, we obtain a similar significance for the $Z' \gamma$ process as in BP1.
However, the significance of the $Z' W$ process is enhanced by the purely left-handed couplings to also have similar values as $Z' \gamma$ for the entire mass range.
The plot shows that ${\cal S}>2$ (${\cal S}>5$) in the entire mass range ($m_{Z'}<200$ GeV and $m_{Z'} > 250$ GeV) in the 8-TeV case for both $Z'\gamma$ and $Z' W$ processes. 
The $Z' Z$ process gives a smaller value of ${\cal S}$ that is less than 2 (5) in the entire mass range for the 8-TeV (14-TeV) case.

\begin{figure}[t]
\begin{center}
\includegraphics[width=75mm]{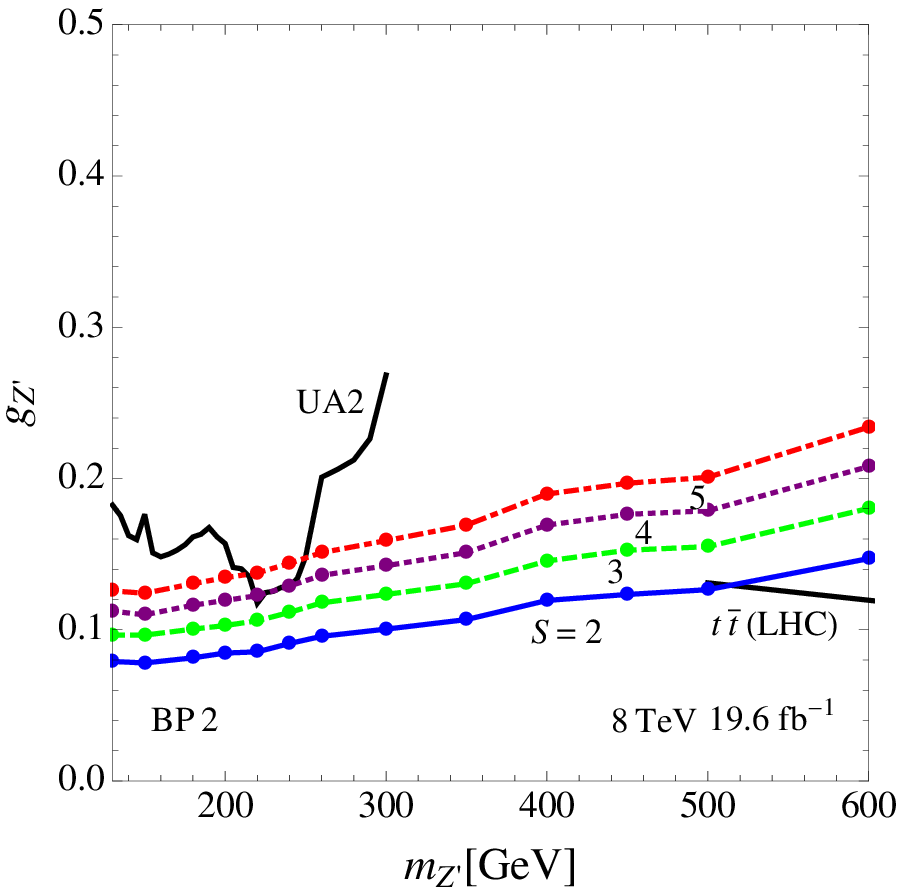}
\hspace{0.5cm}
\includegraphics[width=75mm]{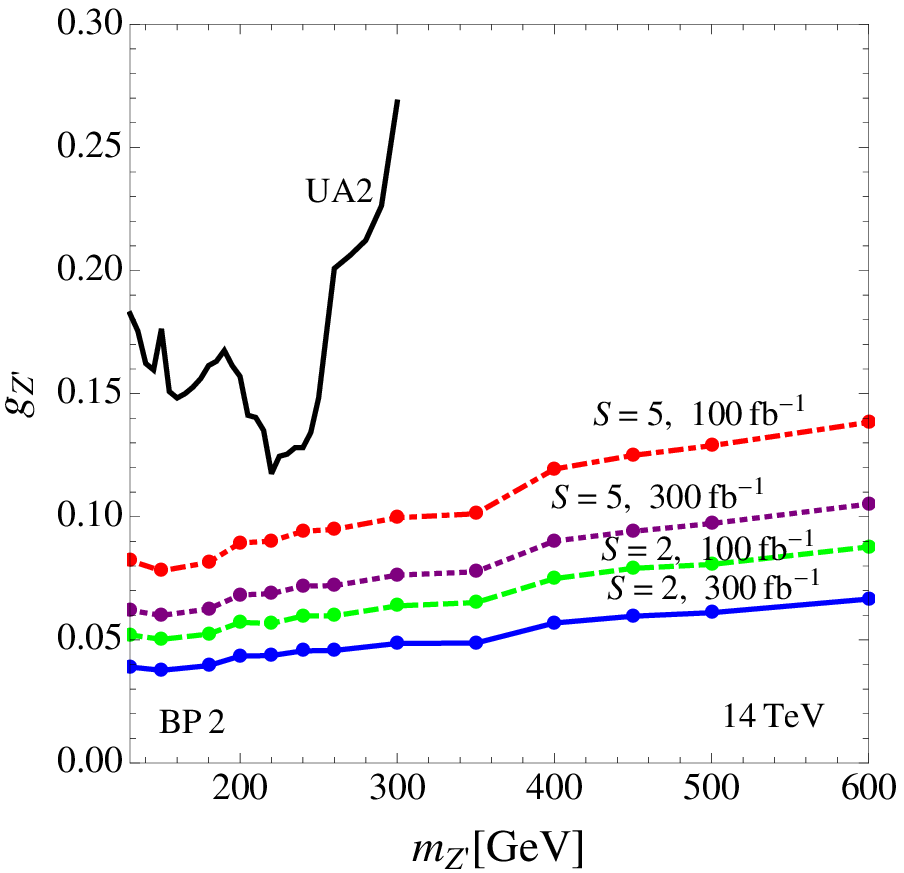}
\hspace{0.5cm}
\includegraphics[width=75mm]{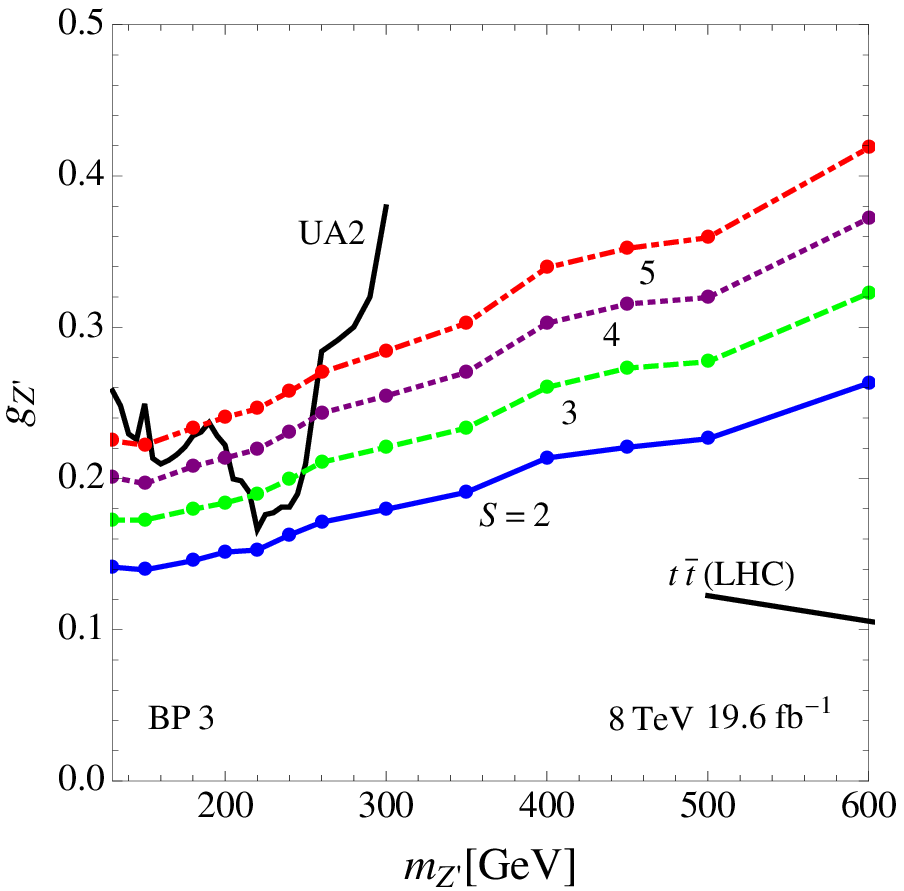}
\hspace{0.5cm}
\includegraphics[width=75mm]{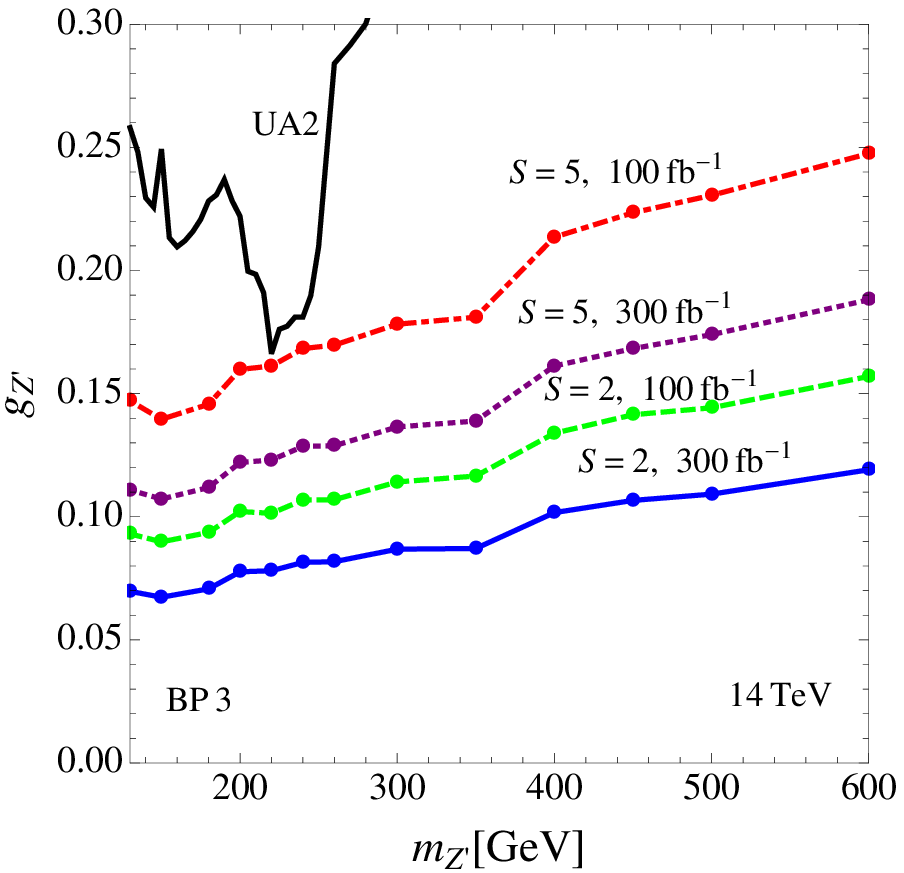}
\end{center}
\caption{Same as Fig.~\ref{const_1h}, but for BP2 (upper plots) and BP3 (lower plots).
}
\label{Sig2}
\end{figure}

Finally, following the analysis for $Z' \gamma$ in the one-Higgs doublet case, we derive here the required $Z'$ gauge coupling $ g_{Z'}^{}$ to achieve specific values of ${\cal S}$ as a function of $m_{Z'}$.
Here we take $K=1.4$ for the backgrounds as in the one Higgs-doublet case.
The upper left (lower left) plot in Figs.~\ref{Sig2} shows the contours of required $g_{Z'}^{}$ for ${\cal S}=2$, 3, 4 and 5 at the 8-TeV LHC with the integrated luminosity of 19.6 fb$^{-1}$ in the scenario of BP2 (BP3). 
The upper limits from the UA2 experiment and the CMS $t \bar t$ data are also shown in these plots. 
For BP2, we find that one can get a stronger constraint on $ g_{Z'}^{}$ as compared to the one-Higgs doublet case in Fig.~\ref{const_1h}.
For BP3, the constraint is weaker but the relative strength between $ g_{Z'}$ values of UA2 limit and that on fixed-significance contours are similar to the one-Higgs doublet case (BP1).
This is because BP3 produces a ratio of the cross sections of $p \bar p \to j j$ and $p p \to \gamma Z'(\to b \bar b)$ that is similar to the one-Higgs doublet case.
The upper right (lower right) plot in Figs.~\ref{Sig2} shows the required $g_{Z'}^{}$ 
for ${\cal S}=2$ and 5 for BP2 (BP3), assuming the 14-TeV LHC with the integrated luminosities of 100 fb$^{-1}$ and 300 fb$^{-1}$.
The 95\% CL upper limit (${\cal S}=2$) is well below the limit from the UA2 experiment, and is about 0.05 (0.04) and 0.08 (0.07) for BP2 and BP3, respectively, at $m_{Z'}=150$ GeV and assuming the integrated luminosity of 100 (300) fb$^{-1}$. 
The minimum value of $g_{Z'}^{}$ for the discovery (${\cal S}=5$) is also below the UA2 limit, which is about 0.08 (0.06) and 0.14 (0.11) for BP2 and BP3, respectively, at $m_{Z'}=150$ GeV and assuming the integrated luminosity of 100 (300) fb$^{-1}$.

\section{Conclusions}

We have studied the collider phenomenology of 
the leptophobic $Z'$ boson from an extra $U(1)'$ gauge symmetry in models with multiple Higgs doublet fields
under the conditions of
(I) no mixing with the Z boson, (I\hspace{-0.5mm}I) no interactions with charged leptons, and (I\hspace{-0.5mm}I\hspace{-0.5mm}I) no FCNC via the $Z'$ mediations. 
We have shown that in the case of $N=1$, all the $\bar{\cal Q}$ charges for quarks have to be the same; {\it i.e.}, $\bar{\cal Q}(Q_L) = \bar{\cal Q}(u_R) = \bar{\cal Q}(d_R)$.  This consequence is relaxed in models with $N > 2$. 
We have explicitly shown that the three conditions cannot be simultaneously met in the case of $N=2$. 

We have discussed the constraint on the $U(1)'$ gauge coupling constant from currently available data of the UA2 experiment and the CMS $pp \to t \bar t$ measurements at the LHC. 
The upper limit on the gauge coupling constant is derived for $m_{Z'} < 300$ GeV from the former and for $m_{Z'} > 500$ GeV from the latter.

We have studied the $pp\to Z'V\to b\bar{b}V$ processes ($V=\gamma,~Z$ and $W^\pm$) with the leptonic decays of $Z$ and $W^\pm$ at the LHC for the $N=1$ and $N \geq 3$ cases. 
We propose that in the case of $N=1$ and $m_{Z'}$ less than the $t\bar{t}$ threshold, 
the $Z'\gamma$ process serves as the most promising discovery mode or provides the most stringent constraint on the $U(1)'$ gauge coupling constant, stronger than that from the UA2 experiment, at the collision energy and integrated luminosity of 8 TeV and 19.6 fb$^{-1}$, respectively.  The $t \bar t$ mode is still the best search channel above the $t\bar{t}$ threshold.

When $N\geq 3$, we have considered four benchmark points (BP1, BP2, BP3 and BP4) for the $Z'$ couplings with quarks.  The benchmark point BP1 exactly corresponds to the $N=1$ case.  In BP2, the couplings are chosen such that the $Z' \to b \bar b$ has the largest branching ratio.  It has a slightly stronger constraint on the gauge coupling from the $Z'\gamma$ process than that in the $N=1$ case. 
We have also found that the benchmark point with purely left-handed couplings (BP3) 
gives similar values of significances for the $Z' \gamma$ and $Z' W $ processes in contrast to the others.  This example shows that a careful comparison between the cross sections of $Z' \gamma$ and $Z' W$ processes will be crucial to reveal the nature of $Z'$ couplings with quarks.  The scenario of BP4 has nonzero $Z'$ couplings only for the right-handed down-type quarks.  This results in null cross section for the $Z'W$ process.  Nevertheless, the $Z'\gamma$ cross section approaches that in BP2 because of the larger gauge coupling strength allowed by UA2.
Finally, we have computed the discovery reach of such a $Z'$ boson at the 14-TeV LHC in both $N=1$ and $N\geq 3$ cases.  
For BP1, the expected $2\sigma$ upper limit on the $Z'$ coupling $g_{Z'}^{}$ has been obtained to be about 0.049~(0.084) for $m_{Z'}^{}=130~(600)$ GeV assuming the integrated luminosity of 300 fb$^{-1}$. 
In addition, the required $Z'$ coupling to reach the $5\sigma$ discovery has been found to be 0.078 (0.13)  for $m_{Z'}^{}=130~(600)$ GeV.

\section*{Acknowledgments}

This research was supported in part by the Ministry of Science and Technology of R.~O.~C. under Grant 
Nos.~MOST-100-2628-M-008-003-MY4, MOST-101-2811-M-008-014, and MOST-103-2811-M-006-030.
C.~W.~C. would like to thank the hospitality of the Kobayashi-Maskawa Institute at Nagoya University where part this work was done during his sabbatical visit.  K.~Y. is supported in part by the JSPS postdoctoral fellowships for research abroad.

\end{document}